\documentclass[aps,prb,twocolumn,groupedaddress,longbibliography]{revtex4-2}%

\usepackage{amsmath}
\usepackage{amsfonts}
\usepackage{amssymb}
\usepackage{graphicx}
\usepackage{xcolor}
\usepackage{lineno}  %
\usepackage{subcaption}
\usepackage{comment}
\usepackage[normalem]{ulem}

\begin{document}


\title{Possibility of Potts Spin-Ice in Rare Earth Twisted-Kagome Layers: an Example of DyAgGe}

\author{Gunnar F. Schwertfeger}
\email{Corresponding author: gschwert@gmu.edu}
\affiliation{Department of Physics and Astronomy, George Mason University, Fairfax, VA
22030, USA}
\affiliation{Quantum Science and Engineering Center, George Mason University, Fairfax, VA
22030, USA}

\author{Po-Hao Chang}
\affiliation{Department of Physics and Astronomy, George Mason University, Fairfax, VA
22030, USA}
\affiliation{Quantum Science and Engineering Center, George Mason University, Fairfax, VA
22030, USA}

\author{Predrag Nikoli\'{c}}
\affiliation{Department of Physics and Astronomy, George Mason University, Fairfax, VA
22030, USA}
\affiliation{Quantum Science and Engineering Center, George Mason University, Fairfax, VA
22030, USA}

\author{Igor I. Mazin}
\affiliation{Department of Physics and Astronomy, George Mason University, Fairfax, VA
22030, USA}
\affiliation{Quantum Science and Engineering Center, George Mason University, Fairfax, VA
22030, USA}

\author{Hari Bhandari}
\affiliation{Department of Physics and Astronomy, University of Notre Dame, Notre Dame, IN 46556, USA}
\affiliation{Stravropoulos Center for Complex Quantum Matter, University of Notre Dame, Notre Dame, IN 46556, USA}
\affiliation{Department of Physics and Astronomy, George Mason University, Fairfax, VA 22030, USA}

\author{Nirmal J. Ghimire}
\affiliation{Department of Physics and Astronomy, University of Notre Dame, Notre Dame, IN 46556, USA}
\affiliation{Stravropoulos Center for Complex Quantum Matter, University of Notre Dame, Notre Dame, IN 46556, USA}

\begin{abstract}

Three-dimensional (3D) spin-ice compounds, predominantly based on the pyrochlore or spinel structure, have been attracting considerable interest in the last two decades. A two-dimensional (2D) analogue is the spin-ice consisting of kagome planes. So far, only Ising spin ice has been discovered in any material, and only one 2D material, the kagome compound HoAgGe, has been identified as a spin ice. In this Letter, we show that another rare-earth twisted Kagome magnetic compound, DyAgGe, is in fact a {\it Potts} spin ice (or color ice), where at each site not two but four spin directions are allowed. We argue that an 8-state spin ice may also exist in this class of systems. We present a first-principles calculation of magnetic interactions in DyAgGe, and confirm the 4-state Potts structure of this material by matching our model predictions with the available experimental data.

\end{abstract}
\maketitle

\paragraph{Introduction:}

Spin ice materials (SIM) are geometrically frustrated magnets in which single-site anisotropy constrains the magnetic moments to lie along a given local "easy" direction\cite{Bramwell2001,Castelnovo2012}. The only known 3D SIMs are pyrochlore-lattice materials with magnetic ions arranged in corner-sharing tetrahedra and the easy direction pointing toward the center of a tetrahedron\cite{Harris1997}. They have been dubbed "spin ice" because their lowest-energy two-in-two-out spin configurations on a tetrahedron \cite{Greedan2001} are equivalent to those of water molecules in ice\cite{Pauling1935}. The extensive geometric frustration of pyrochlore spin ice leads to rich and unique magnetic properties as well as exotic states of matter featuring fractionalization\cite{Hermele2004a, Castelnovo2012}, Dirac strings and magnetic monopoles\cite{Bramwell2009, Morris2009, Castelnovo2008, Armitage2014}.

\begin{figure}[h!]
    \includegraphics[width=0.98\linewidth]{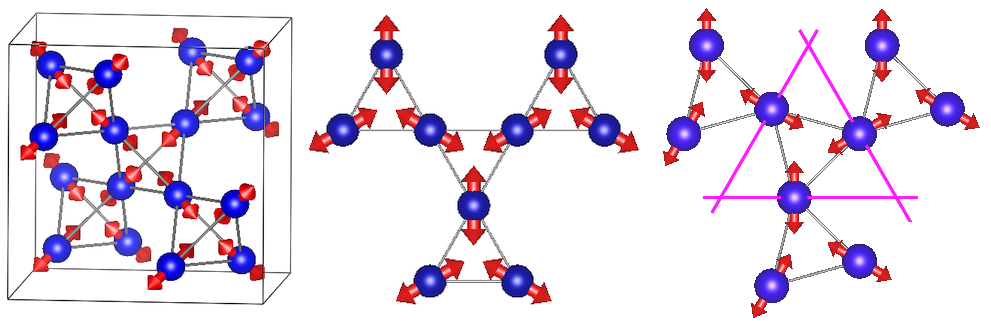}
    \caption{Schematics of, from left to right:  a pyrochlore 3D SIM; a 2D kagome SIM; a twisted kagome SIM, as in HoAgGe. Magenta lines indicated mirror planes.
    }
    \label{SIM}
\end{figure}

The 2D variant of spin ice on the kagome lattice, (KL), was studied theoretically \cite{Lacroix2002, Tchernyshyov2011, Tchernyshyov2012}, and some progress was made in creating an artificial spin ice \cite{Schiffer2006, Cumings2011}. However, only very recently has a spin ice been identified in a natural kagome material HoAgGe \cite{Gegenwart}. The kagome lattice consists of corner-sharing triangles, and its ice rule is that the spins on the triangle vertices point two-in-one-out or one-in-two-out, see Fig.\ref{SIM}. A KL can be distorted by twisting without reducing the degree of frustration, as in the HoAgGe compound \cite{Gegenwart} shown in Fig. \ref{SIM}. In that case, the local (site-)symmetry is orthorhombic, and, specifically in HoAgGe, the easy direction is perpendicular to the local mirror plane. 

All experimentally reported SIM so far, in 2D or 3D, have exclusively Ising spins. Generalizations to multi-state degrees of freedom, sometimes dubbed "color ice" have appeared only in theoretical studies \cite{bionic, tripole, Wu2014, Gingras2016, Moessner2025, Moessner2026} that do not address realistic materials or experiments. For instance, a 4-spin generalization of pyrochlores, considered in Ref. \cite{bionic}, postulates that each pyrochlore site has 4 admissible spin directions, ($\bar{1},1,1),(1,\bar{1},1),(1,1,\bar{1}),(\bar{1},\bar{1},\bar{1})$, even though this is not compatible with the trigonal site symmetry of magnetic sites. 

Naively, one may think that the easy direction must coincide with one of the local high-symmetry directions (note that the model of Ref. \cite{bionic} violates this rule), i.e., in the twisted Kagome lattice, (TKL), one of the three local orthorhombic directions.
This intuition has its origin in an implicit assumption that magnetic anisotropy energy can be expanded in powers of the ratio of the spin-orbit coupling strength and a characteristic crystal field splitting, $\xi_{\rm SOC}/\Delta_{\rm CF}\ll 1$. This assumption is correct for 3d and 4d metals, and sometimes even for 5d, but is completely broken in f-elements. In fact, in most cases the opposite is true, $\Delta_{\rm CF}/\xi_{\rm SOC}\ll 1 \quad$\footnote{Since SOC in 4f elements is in hundreds of meV, an order of magnitude stronger that in 3d metals, and the Coulomb crystal field, due to the small size of the f-shell, is suppressed by screening by outer shells and the ligand field is suppressed due to small hoping}, and a new perturbation theory in this parameter can be worked out\cite{Igor2023,Lee2025}, which expresses magnetic anisotropy in terms of Wigner coefficients, and all terms up to the sixth order in directional cosines are formally of the same order. As a result, the easy axis is no longer tied to a high-symmetry direction, but can be completely arbitrary, depending of the crystal field splittings and f-shell filling. For example, in site-hexagonal kagome metals EuMn$_6$Sn$_6$ and  HoMn$_6$Sn$_6$ this expansion generates the so-called easy-cone anisotropy with a cone angles of 40--50$^\circ$, as confirmed by the experiment\cite{Igor2023}. 

\begin{figure}[h!]
    \includegraphics[width=0.98\linewidth]{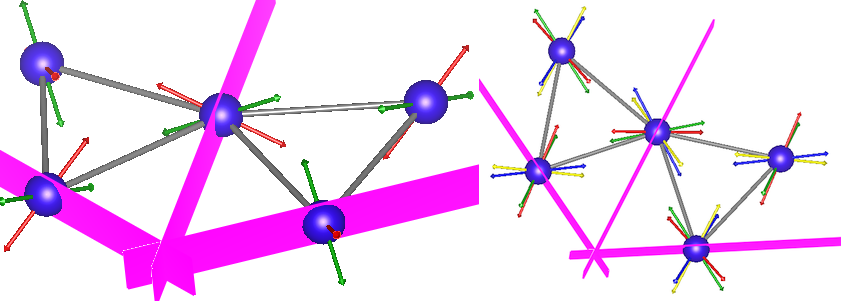}
    \caption{Schematics of, from left to right:  a twisted kagome SIM with a tilted easy axis, as in DyAgGe; a twisted kagome SIM with a most general easy axis; the same, for a perfect kagome. Magenta lines indicated mirror planes. 
    }
    \label{Potts}
\end{figure}

For an orthorhombic site symmetry all 7 f-levels are split, thus providing an immense trove of possibilities, and it is merely fortuitous that in HoAgGe the theory predicts\cite{Gunnar_HoAgGe}, and the experiment confirms, a high-symmetry easy axis. At the same time, nothing prevents the same TKL with a different rare earth (different f-occupation) or different ligands (different crystal field) to have an easy direction anywhere, including deviation from a pure in-plane direction and/or from a high-symmetry direction in the plane. In the former case we will now have four possible spin states (see Fig. \ref{Potts}), i.e., instead of an Ising SIM we have a {\it 4-state} Potts SIM. In the latter case we would have eight states, despite the site symmetry being only $m2m$, and not $mmm$, as in the perfect kagome. In that case, the local spin would be described by an {\it 8-states} Potts model. 

As mentioned, all experimentally reported SIMs so far belong to the Ising class. Numerous theoretical excursions into possible generalizations onto the Potts realm (see Ref. \cite{Moessner2026} and references therein) have not even attempted to relate the concept with any actual material.
In this Letter we argue that the well-known family of XYZ twisted kagome materials, where X is and f-metal and Y an Z can be literally anything, from Li to Au, with over 50 known members\cite{Huang2023}, and definitely dozens of unknown, is a fertile ground for 4-- and 8--states Potts spin ices, and demonstrate one such material, DyAgGe, which appears to belong to the former class.

\paragraph{Magnetic Hamiltonian for twisted kagome: }
For a TKL of $N$ sites and with local magnetic moments, $M = \{ \hat{\bf S}_0, \hat{\bf S}_1, ... \hat{\bf S}_{N-1} \}$, we define the essential Hamiltonian, Eq.\ref{eq:hamiltonian_ij}, where $J_{ij}$ is the Heisenberg exchange parameter between pairs of Dy atoms $i$, $j$, and $h$ is the applied magnetic field along $\hat{\mathbf h}$. 

\begin{equation}
H[M;J_{ij},h] = -\sum_{i,j} J_{ij} \left(\hat{\bf S}_i\cdot\hat{\bf S}_j\right) - h \sum_i \hat{\bf S}_i \cdot \hat{\bf h} \label{eq:hamiltonian_ij}
\end{equation}

\begin{figure}[h!]
\begin{center}
    \includegraphics[scale=0.3]{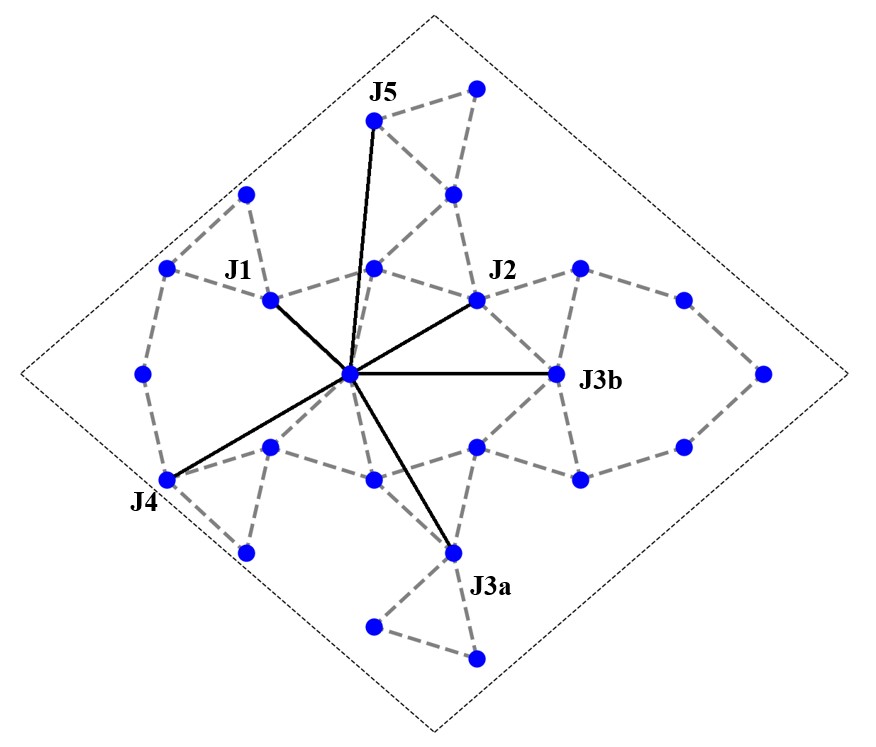}
    \caption{$J_1$ to $J_5$ Dy-Dy interactions depicted for a single atom.}
    \label{fig:ref_bonds}
\end{center}
\end{figure}

Combining the interactions 
as defined in Fig.\ref{fig:ref_bonds} (note that third nearest neighbor bonds must be split, by symmetry, into two different types, even if the same length), we get:

\begin{equation}
H[M;J_\alpha,h] = -\sum_\alpha J_{\alpha} \sum_{\left< i, j\right>_\alpha} \hat{\bf S}_i\cdot\hat{\bf S}_j - h \sum_i \hat{\bf S}_i \cdot \hat{\bf h} \label{eq:hamiltonian_alpha}
\end{equation}

\paragraph{Density functional calculations of the magnetic Hamiltonian}

We calculated the exchange parameters, $J$, perturbatively using the the Green's function method \cite{antropovExchangeInteractionsMagnets1997,katsnelsonFirstprinciplesCalculationsMagnetic2000} as implemented in OpenMX 3.9 code\cite{terasawaEfficientAlgorithmBased2019,omxgf2004} after assuring that the f-orbital occupation respects the second Hund's rule. Previuosly, this methodology was succesfully applied to the sister compound HoAgGe \cite{Gunnar_HoAgGe,Bhandari2025}; note that the dominant coupling $J_{1}$ is ferromagnetic in HoAgGe, whereas it is antiferromagnetic in DyAgGe, and the easy axis in the latter is tilted away from the plane. Technical details can be found in the Supplementary Information.

\begin{table}[h]
\begin{center}
\caption{
    Heisenberg exchange parameters for Dy-Dy interactions in the twisted kagome lattice up to 5th nearest neighbor calculated using a DFT and GF method. The values are normalized to $|J_1|=2.294$ meV. Negative values are antiferromagnetic}
    \label{table:Dy_params}
    \renewcommand*{\arraystretch}{1.4}
    \setlength{\tabcolsep}{7.0pt}
    \begin{tabular}{c|cccccc}
    \hline
        $J_{\alpha}$ & $J_{1}$ & $J_{2}$ & $J_{3a}$ & $J_{3b}$ & $J_{4}$ & $J_{5}$ \tabularnewline
    \hline
        $J_{\alpha}/|J_{1}|$ & -1.00 & -0.28 & -0.15 & -0.24 & -0.18 & -0.16 \tabularnewline
    \hline
    \end{tabular}
\end{center}
\end{table}

\begin{figure}[h!]
    \includegraphics[scale=0.28]{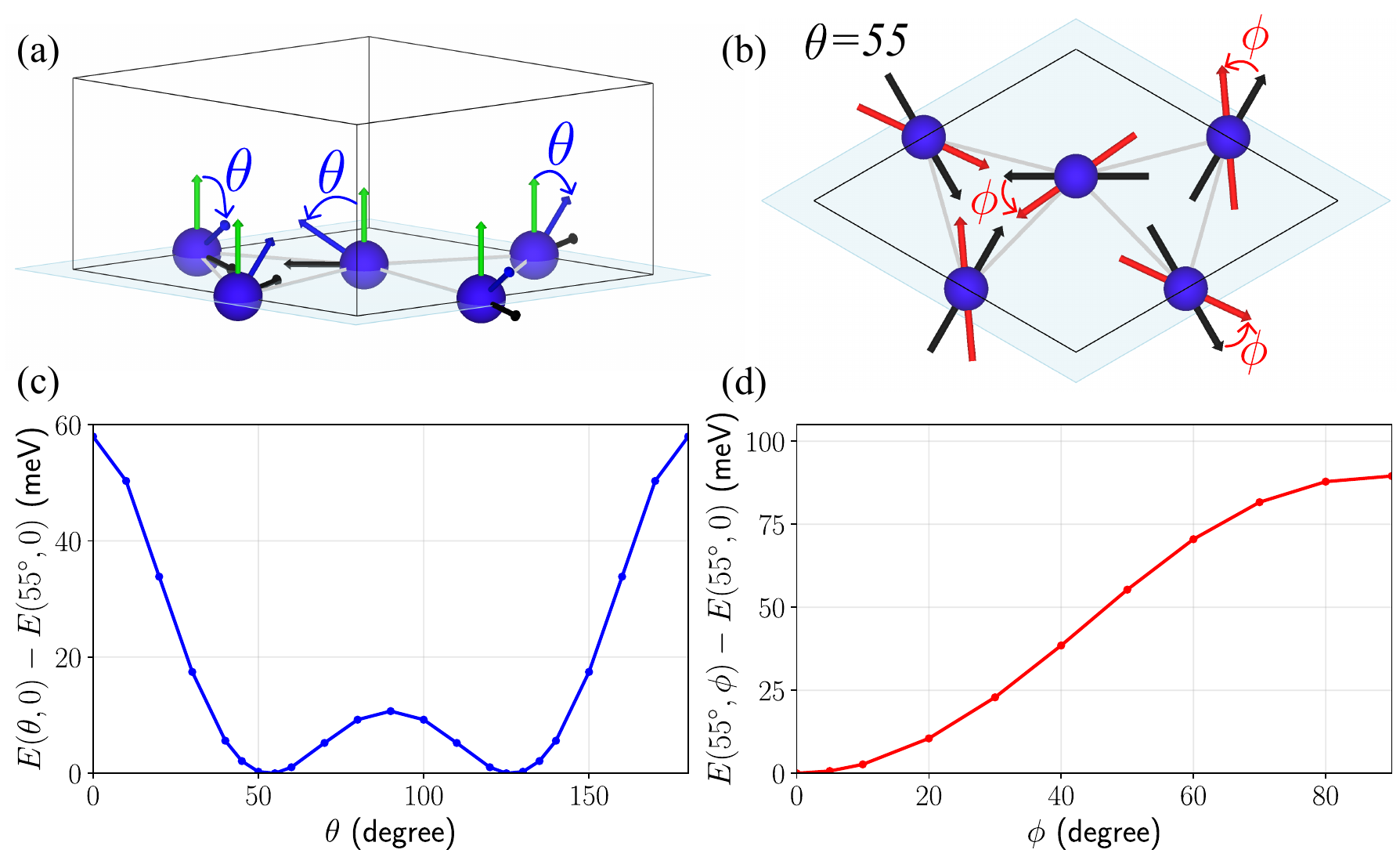}
    \caption{Calculation of the easy axis for the Dy in DyAgGe lattice. First, we fixed the azimuthal angle $\phi$ to zero (from the mirror plane normal) and varied the polar angle $\theta$ (a,c). We found the minimum to be at $\theta\pm 55^\circ$. Then, keeping $\theta=55^\circ$, we varied $\phi$ (b,d), and found that $\theta=0$ was the minimum.
    The two distinct minima seen in (c) lead to the four-state Potts model (a) and (b).
    }
    \label{fig:Dy_angle}
\end{figure}

The easy axis was determined through evaluation of the magnetic anisotropy as a function of angular dependence, in two steps, as described in Fig. \ref{fig:Dy_angle}. We find that the easy axis corresponds to $\theta\approx 55^\circ,\ \phi=0$, indicating the exciting possibility of DyAgGe being the first identified Potts Spin-Ice.

\paragraph{Experimental results} 
Samples of DyAgGe were grown by the flux method using the recipe provided in \cite{morosan2004thermodynamic}. Magnetization measurements for B $||$ [100], [120], and [001] ($\hat{\mathbf x}$, $\hat{\mathbf y}$, and $\hat{\mathbf z}$ respectively) are consistent with those observed  in \cite{morosan2004thermodynamic}. Further details of the experimental procedures are provided in the Supplementary materials. 
The preliminary magnetic structure of polycrystalline samples of DyAgGe, HoAgGe, and TbAgGe through neutron diffraction measurements have been reported in \cite {baran1998magnetic}, where the magnetic structure of DyAgGe is reported to be a square modulated magnetic structure with Dy magnetic moments lying in the b-c plane making an angle of 50(7) degree with the c-axis, which is supported by the observation of magnetization plateau observed for both in-plane and out of plane magnetization measurements.

\paragraph{Discussion of results:}

\begin{figure}[h!]
\begin{center}
    \includegraphics[scale=0.25]{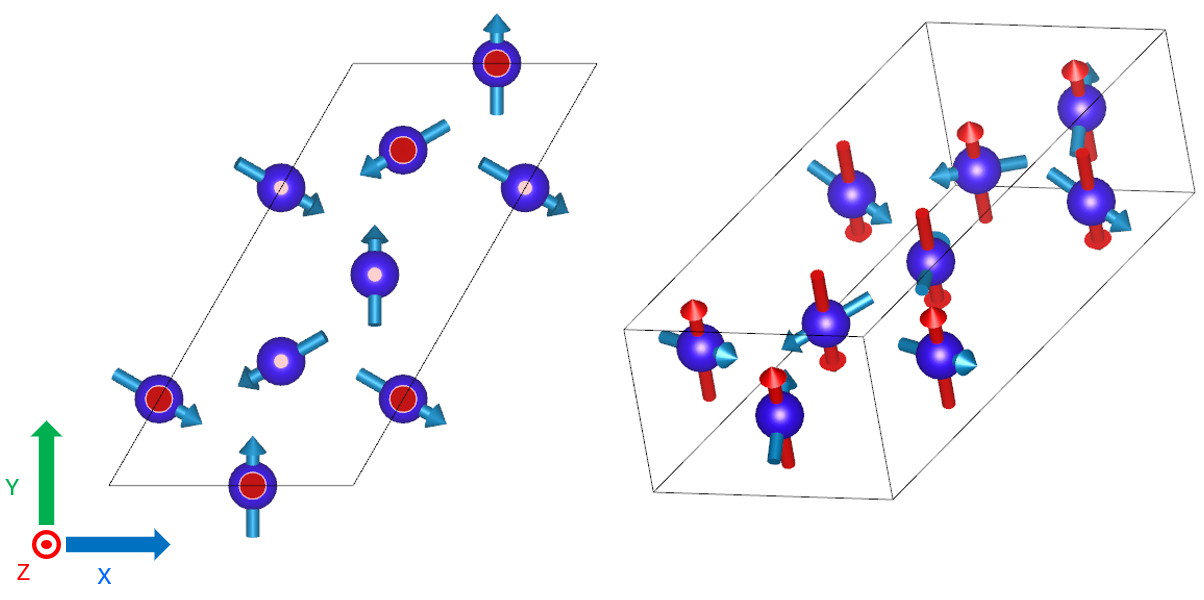}
    \caption{Theorized ground state orientation of the Dy atoms with MSG: Pn'C2'.
    Out-of-plane magnetic moment components ($S_i^z$) visualized using \textcolor{red}{red} arrows for ease of viewing.
    }
    \label{fig:ground_state}
\end{center}
\end{figure}

In the sister compound, HoAgGe, we successfully described  $M(h)$ for the two relevant field directions using a Reduced-Configuration-Space search (RCS) method\cite{Gunnar_HoAgGe} and the DFT exchange parameters, much better that the empirical parameters used in Ref. \cite{Gegenwart}. In particular, we were able to identify two unexplained intermediate phases\cite{Gunnar_HoAgGe}.
Now, using this method\footnotemark[898] we constructed the  magnetization curve $M(h)$ for DyAgGe for $\hat{\mathbf x}$, $\hat{\mathbf y}$, and $\hat{\mathbf z}$ directions (see Fig. \ref{fig:ground_state} for the global frame definition).

\footnotetext[898]{Additional details on adapting the Potts Hamiltonian to the RCS analysis model is provided in the supplemental materials.}

Using our experimental measurements, we can compare the results of our computational analysis using the DFT calculated exchange parameters.

All major and minor magnetization plateaus seen in experiment are reproduced. Due to the extreme hysteresis present in the experimental measurements, we have some freedom to normalize our results with the crystal saturation for comparison. We found that after roughly normalizing to the lowermost saturation limit that our theorized plateaus matched well with those experimentally observed. There is room for slight variation of exchange parameters to further affect fitting, but through inspection of various parameter phase diagrams\footnotemark[141] we found this to be of little importance as the initial calculated parameters reproduced the results sufficiently.

\footnotetext[141]{Phase diagrams of varied exchange parameters as an external field is applied in the $\hat{\mathbf x}$, $\hat{\mathbf y}$, and $\hat{\mathbf z}$ directions are included in the Supplemental Materials.}

Through the use of an RCS Search method we arrive at a potential ground state shown in Fig.\ref{fig:ground_state}. This state has zero magnetization and is of the Pn'C2' magnetic space group. Due to the large hysteresis observed in the crystal magnetization near the ground state we suspect that this state may not actually be observed experimentally. Unsurprisingly, there are a great many degenerate ground states close in energy near zero field\footnotemark[767]. Because of how near in energy these states are, it is reasonably likely that other outside effects may affect the system causing the crystal to shift between these neighboring states.

\footnotetext[767]{The degeneracy of states is shown directly in the Supplemental Materials}

\begin{figure}[h!]
\begin{center}
    \includegraphics[scale=0.25]{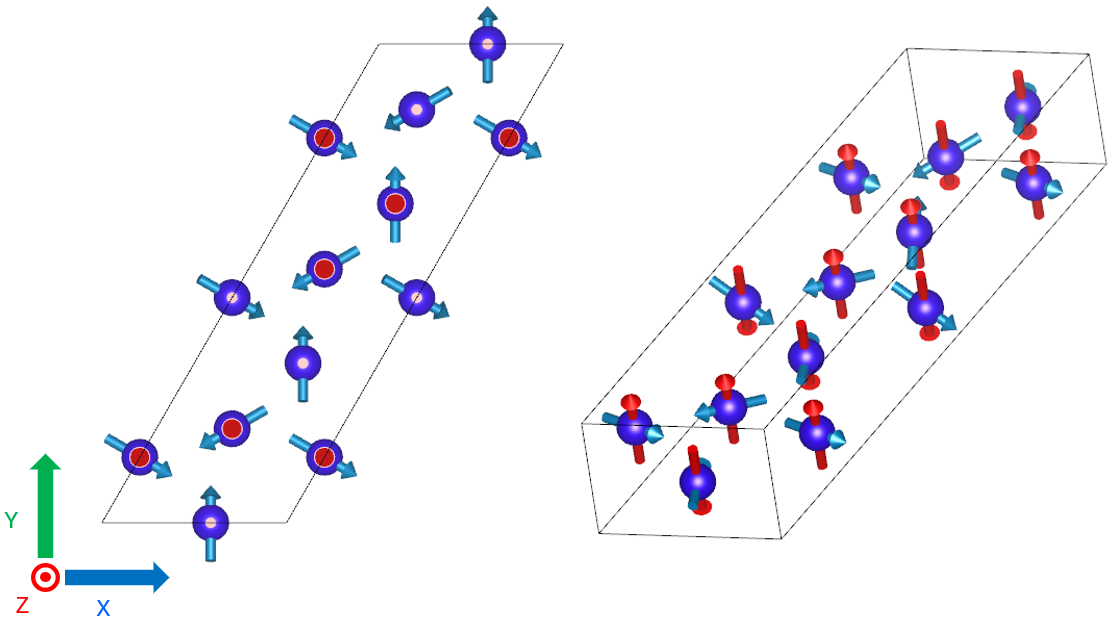}
    \caption{Possible $1/9$th ground state orientation of the Dy atoms with MSG: C2'.
    Out-of-plane magnetic moment components ($S_i^z$) visualized using \textcolor{red}{red} arrows for ease of viewing.
    }
    \label{fig:one_ninth}
\end{center}
\end{figure}

Instead of the zero magnetization state, due to the hysteresis seen in experiment near zero field, a 1/9th magnetization state of magnetic space group C2' may be observed. This state is shown in Fig.\ref{fig:one_ninth}.

\paragraph{Possibility of Additional Interactions:}

While we believe Heisenberg exchange, Dzyaloshinskii-Moriya, and dipole-dipole interactions to dominate this system, there is the possibility that other interactions could affect the system and lift degeneracies between states observed using the RCS method. Two possible interactions to consider are symmetric anisotropic exchange, and biquadratic coupling.

Biquadratic coupling is the only interaction to mix the in-plane and out-of-plane spin components. Thus, its main role is to establish a correlation between these two order patterns, otherwise completely decoupled. Addressing this coupling is an interesting issue, but it dramatically increases the complexity, as now the optimization must be performed in the whole 3D space, and goes beyond the scope of this paper.

\footnotetext[466]{A more detailed explanation on why Biquadratic coupling is incompatible with the implemented model is provided in the Supplemental Materials}

The effects of symmetric anisotropic exchange as they are fully separable, should be absorbable into the corresponding Heisenberg exchange parameters provided the in-plane components are reasonably comparable. For this reason, any affects of SAE would be absorbed and visible in the Heisenberg parameter phase diagrams (shown in supplemental materials). As no obvious minor adjustments to the Heisenberg exchange parameters affect our fitting meaningfully, we omit SAE.

The exclusion of these interactions did not appear to affect our results substantially as we still reproduce the experiment sufficiently.

\paragraph{Confirmation with Monte Carlo Simulations:}

\begin{figure*}[t]
\includegraphics[scale=0.5]{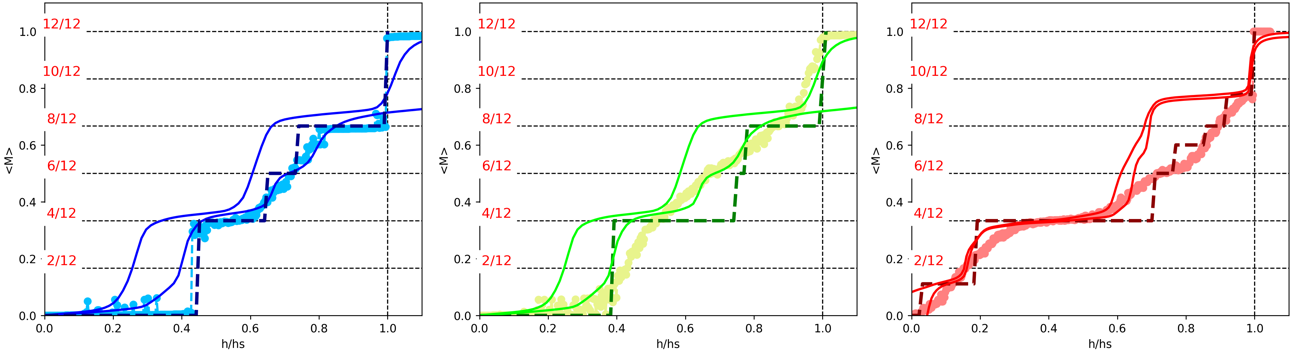}
\centering
\caption{Average magnetization for the Dy atoms as an external magnetic field is applied in the (from left to right) \textcolor{blue}{X}, \textcolor{green}{Y}, and \textcolor{red}{Z} directions. The solid lines represent the experimentally measured magnetization, light dots indicate the magnetization found via MH Monte Carlo, and the dark dashed lines representing magnetization found using an RCS Search.}
\label{fig:magnetization}
\end{figure*}

Using the calculated exchange parameters we also conduct Monte Carlo simulations of our system in an attempt to confirm our analysis. All Monte Carlo simulations utilized a standard Metropolis-Hastings algorithm with temperature near absolute zero. The results of the simulations agree with the RCS analysis and are shown in Fig.\ref{fig:magnetization}.

Again due to the extreme hysteresis present in the magnetization curves observed experimentally, we have some liberty in scaling the Monte Carlo results to the saturation limit of the crystal. We again find rough agreement with both the experimental observations and RCS predictions further verifying our analysis.

\paragraph{Temperature Considerations:}

As the Magnetic Ordering Temperature of DyAgGe has been experimentally measured to be in the range of $T=5-12$K, we are able to ignore the affects of interactions with average energy sufficiently below the temperature threshold of approximately $~1$meV (less than half the strength of the nearest-neighbor $J_1$ interaction). As the furthest neighbor interaction considered was the fifth-nearest-neighbor, $J_5$, we consider anything below its interaction strength of $0.374$ meV as sufficiently small and able to be safely excluded from our analysis.

We attribute the premature shift to the $7/9$ state from the $1/3$ state seen in the experiment when the external field is applied along $\hat{\mathbf z}$ (Fig.\ref{fig:magnetization} {\it right}) to these temperature effects. The energy difference between these states is well within this temperature limit and for this reason the interstitial magnetic states ($1/2$, $3/5$, and $2/3$) may not be easily observed. This is shown more directly in the supplemental materials.

\paragraph{Conclusions:}

Through the use of a DFT+GF method to calculate Heisenberg Exchange parameters, we were able to reproduce many of the experimentally observed results using an RCS Search method. These results were also replicated using a standard Monte-Carlo Metropolis-Hastings algorithm.

With the reproduction of the experimental measurements, we are confident in our theoretical model of the DyAgGe crystal being at least a four-state Potts model and the first to be identified in its new class of Potts Spin Ice Materials. We predict that other cousin materials containing similar rare-earth atoms may host even more exotic spin structures such as the possible eight-state Potts Spin Ice.

\begin{acknowledgments}
This work was supported by the National Science Foundation under Award No. DMR-2403804.

\end{acknowledgments}

\bibliography{bib.bib, dft.bib, pn.bib}

\clearpage
\onecolumngrid
\appendix

\setcounter{equation}{0}
\setcounter{figure}{0}
\setcounter{table}{0}
\setcounter{section}{0}
\renewcommand{\theequation}{S\arabic{equation}}
\renewcommand{\thefigure}{S\arabic{figure}}
\renewcommand{\thetable}{S\arabic{table}}

\begin{center}
\textbf{\large Supplemental Material}
\end{center}

\section{Defining the Potts Hamiltonian}

The Potts Hamiltonian is defined:

\begin{equation}
H[M;J_{ij},h] = -\sum_{i,j} J_{ij} \left(\hat{\bf S}_i\cdot\hat{\bf S}_j\right) - h \sum_i \hat{\bf S}_i \cdot \hat{\bf h} \label{sup:eq:hamiltonian_ij}
\end{equation}

\begin{figure}[h!]
\begin{center}
    \includegraphics[scale=0.3]{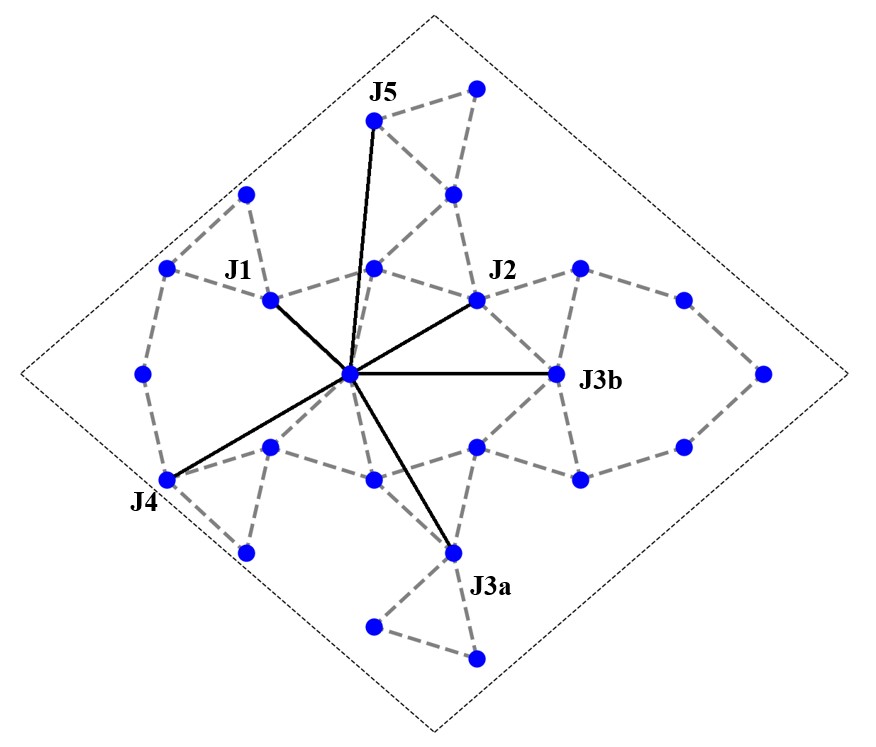}
    \caption{$J_1$ to $J_5$ Dy-Dy interactions depicted for a single atom.}
    \label{sup:fig:ref_bonds}
\end{center}
\end{figure}

We group Dy-Dy interactions by symmetry and denote them using $\alpha$. The resulting Hamiltonian after grouping these interactions is Eq. \ref{sup:eq:hamiltonian_alpha} and the Dy-Dy interactions are shown in Fig \ref{sup:fig:ref_bonds}.

\begin{equation}
H[M;J_\alpha,h] = -\sum_\alpha J_{\alpha} \sum_{\left< i, j\right>_\alpha} \hat{\bf S}_i\cdot\hat{\bf S}_j - h \sum_i \hat{\bf S}_i \cdot \hat{\bf h} \label{sup:eq:hamiltonian_alpha}
\end{equation}

The Dy easy axis are orientated with respect to the directions defined in the main paper.

\section{DFT Calculations}

Density functional theory (DFT) calculations were performed using
the linear combination of numerical atomic orbitals (NAO) method implemented
in the OpenMX code \cite{OpenMX,ozakiVariationallyOptimizedAtomic2003}.
The core electrons were described using norm-conserving pseudopotentials
\cite{morrisonNonlocalHermitianNormconserving1993,vanderbiltSoftSelfconsistentPseudopotentials1990}.
Sufficiently large NAO basis sets were selected as Dy8.0-$s3p3d2f1$,
Ag7.0-$s3p2d2$, and Ge7.0-$s3p2d2$. Exchange-correlation effects
were treated using the Perdew-Burke-Ernzerhof (PBE) generalized gradient
approximation \cite{perdewGeneralizedGradientApproximation1996}.
To account for the strong electron correlations and localized magnetic
moments of the Dy $4f$ states, a Hubbard $U$ correction was included
using the fully localized limit double-counting scheme \cite{liechtensteinDensityfunctionalTheoryStrong1995,dudarevElectronenergylossSpectraStructural1998}
with an effective parameter of $U-J=8\text{ eV}$. Brillouin zone
integration was performed using a $5\times6\times9$ $k$-point mesh,
which ensured that magnetic anisotropy energy differences converged
to within 1 meV/Dy. Under these parameters, the calculated orbital
and spin moments at the Dy sites were both consistently $5\text{ }\mu_{\mathrm{B}}$,
satisfying Hund's rules.

Upon achieving charge self-consistency, the magnetic exchange coupling
constants ($J_{\alpha}$) were calculated via the Green's function method
\cite{antropovExchangeInteractionsMagnets1997,katsnelsonFirstprinciplesCalculationsMagnetic2000}
implemented in OpenMX 3.9 \cite{terasawaEfficientAlgorithmBased2019,omxgf2004}. Calculated values of $J_\alpha$ are shown in the main paper.
This approach enables the direct evaluation of pairwise exchange interactions
up to arbitrary interionic distances starting from a single magnetic
configuration. The validity of this framework has been demonstrated
in the isomorphic compound HoAgGe \cite{Gunnar_HoAgGe,Bhandari2025}; notably,
it correctly captures the qualitative distinction in the dominant
nearest-neighbor coupling parameter ($J_{1}$ is ferromagnetic in
HoAgGe, whereas it is antiferromagnetic in DyAgGe). 

\section{Manipulation of Potts Hamiltonian for Analysis}

\subsection{Splitting of the Potts Hamiltonian}

Through simple algebraic manipulation of Eq.\ref{sup:eq:hamiltonian_alpha} it is clear that the quadratic Hamiltonian can be split into an in-plane and out-of-plane system. While this is immediately obvious for a Hamiltonian consisting of only Heisenberg exchange and Zeeman terms, this can also be shown for the Dzyaloshinskii-Moriya Interaction as well as the Dipole-Dipole Interaction. This is argued more thoroughly in later sections.

\begin{figure}[h!]
\begin{center}
    \includegraphics[scale=0.6]{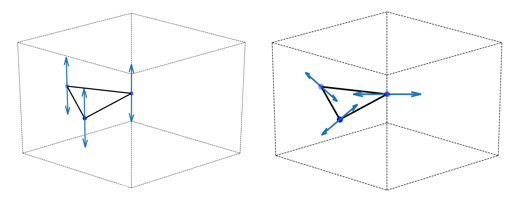}
    \caption{Resulting split Dy spin system. A standard colinear Ising system along $\hat{\mathbf z}$ and a non-colinear system in the $XY$ plane.}
    \label{fig:spin_split}
\end{center}
\end{figure}

For this reason, we can now write our Hamiltonian as Eq.\ref{eq:hamiltonian_split} and split the Dy's four-state Potts model into two simpler and more familiar two-state models (Fig.\ref{fig:spin_split}). 

\begin{equation}
    H[M;J_\alpha,h] = H_{xy}[M;J_\alpha,h] + H_{z}[M;J_\alpha,h]
    \label{eq:hamiltonian_split}
\end{equation}

The two resulting two-state Ising systems are individually much easier to analyze independently than the original four-state Potts system without losing any details of the original crystal Hamiltonian.

\subsection{Dzyaloshinskii-Moriya Interaction}

As the $XY$ plane for the twisted Kagome lattice is a mirror plane, it follows that all DMI vectors must be aligned with the $\hat{\mathbf z}$ direction, perpendicular to the lattice. This immediately allows us to ignore any effects of DMI for the colinear Ising Hamiltonian, $H_z$. Further, as the in-plane components of the Dy atoms magnetic moments are always $2 \pi / 3$ rad or parallel to all others, the in-plane affects of the Dzyaloshinskii-Moriya Interaction may safely be absorbed into the spin exchange parameters, $J_\alpha$. Or shown more explicitly: 

For any two Dy atoms, $i$ and $j$;

\begin{align}
E_{dm} &= \vec{\mathbf D}_{ij} \cdot \left( \hat{\mathbf S}_i \times \hat{\mathbf S}_j \right)\label{eq:dmi}\\
&= \left({\mathbf D}_{ij}\right)^z \left( {\mathbf S}_i^x {\mathbf S}_j^y - {\mathbf S}_i^y {\mathbf S}_j^x \right)\nonumber
\end{align}

As only the in-plane components of the Dy atoms magnetic moments are required to fully explore the DM interaction, the effect may be safely explored purely in the non-colinear in-plane model and ignored entirely in the colinear out-of-plane model.

Further, as the directions of neighboring Dy atoms magnetic moments are well defined (all possible moments always being a multiple of $2 \pi /3$ radians apart), we may make the substitution of Eq.\ref{eq:spin_sub} into Eq.\ref{eq:dmi}:

\begin{equation}
\hat{\mathbf S}_j = \pm \tilde{\mathbf R}\left(2\pi n/3,\hat{\mathbf z}\right) \hat{\mathbf S}_i\label{eq:spin_sub}
\end{equation}

Where $\tilde{\mathbf R}_z\left( \theta ,\hat{\mathbf \omega}\right)$ is a rotational transformation by an angle, $\theta$, about an arbitrary $\hat{\mathbf \omega}$ axis. In this case, a rotation by some multiple of $2 \pi / 3$ about the $\hat{\mathbf z}$ axis.

Giving us:

\begin{align}
E_{dm} &= \left({\mathbf D}_{ij}\right)^z \left( {\mathbf S}_i^x {\mathbf S}_j^y - {\mathbf S}_i^y {\mathbf S}_j^x \right)\nonumber\\
 &= \pm \left({\mathbf D}_{ij}\right)^z \Big( {\mathbf S}_i^x \tilde{\mathbf R}\left(2\pi n/3,\hat{\mathbf z}\right)_k^y {\mathbf S}_i^k \nonumber\\ &\qquad\qquad\qquad - {\mathbf S}_i^y \tilde{\mathbf R}\left(2\pi n/3,\hat{\mathbf z}\right)_k^x {\mathbf S}_i^k \Big)\nonumber\\
&= \Big\{ \pm \frac{\sqrt{3}}{2}\left({\mathbf D}_{ij}\right)^z \left( \left({\mathbf S}_i^x\right)^2 + \left({\mathbf S}_i^y\right)^2 \right),\quad 0 \Big\}\label{eq:dmi_result}
\end{align}

Allowing us to safely absorb the DM interaction into the corresponding exchange interaction parameters, $J_\alpha$. 

\subsection{Dipole Dipole Interaction}

As with the Dzyaloshinskii-Moriya and exchange interactions, the Dipole-Dipole interaction may also be split into in-plane and out-of-plane components through simple algebraic manipulation. This again is a direct consequence of the two-dimensionality of the lattice as for every pair of Dy atoms $i$ and $j$, their displacement perpendicular to the plane is always zero by definition. ($\forall \quad i,j ;\quad \left( \vec{\mathbf r}_j - \vec{\mathbf r}_i \right) \cdot \hat{\mathbf z} = {\mathbf r}_{ij}^z = 0$).

\begin{align}
E_{dd}
&= - \frac{g_{dd}}{|\vec{\mathbf r}_{ij}|^3}\left[ 3\left(\hat{\mathbf S}_i\cdot\hat{\mathbf r}_{ij}\right) \left(\hat{\mathbf S}_j\cdot\hat{\mathbf r}_{ij}\right) -\left(\hat{\mathbf S}_i\cdot\hat{\mathbf S}_j\right) \right] \label{eq:dipole}\\[4pt]
&= - \frac{g_{dd}}{|\vec{\mathbf r}_{ij}|^3} \Bigl[ 3\left(\mathbf S_i^x {\mathbf r}_{ij}^x+\mathbf S_i^y {\mathbf r}_{ij}^y\right) \left(\mathbf S_j^x {\mathbf r}_{ij}^x+\mathbf S_j^y {\mathbf r}_{ij}^y\right) \nonumber\\ &\qquad\qquad\quad -\left(\mathbf S_i^x\mathbf S_j^x +\mathbf S_i^y\mathbf S_j^y\right) \Bigr] \nonumber\\ &\quad +\frac{g_{dd}}{|\vec{\mathbf r}_{ij}|^3} \left(\mathbf S_i^z\mathbf S_j^z\right)\nonumber
\end{align}

Where $g_{dd}$ is defined:
\begin{equation*}
g_{dd} = \frac{\mu_0 \mu_B^2 g^2}{4 \pi}
\end{equation*}

Through grouping we may then simply redefine the Dipole-Dipole interaction in terms of purely in-plane and out-of-plane components:
\begin{align}
\left(E_{dd}\right)_{xy} &= \frac{g_{dd}}{|\vec{\mathbf r}_{ij}|^3}
\Bigl[
\left(\mathbf S_i^x\mathbf S_j^x
+\mathbf S_i^y\mathbf S_j^y\right)
\nonumber\\
&\qquad
- 3\left(\mathbf S_i^x {\mathbf r}_{ij}^x+\mathbf S_i^y {\mathbf r}_{ij}^y\right)
 \left(\mathbf S_j^x {\mathbf r}_{ij}^x+\mathbf S_j^y {\mathbf r}_{ij}^y\right)
\Bigr]\label{eq:dipole_xy}\\
\left(E_{dd}\right)_z &= \frac{g_{dd}}{|\vec{\mathbf r}_{ij}|^3}
\left(\mathbf S_i^z\mathbf S_j^z\right)\label{eq:dipole_z}
\end{align}

Simply leaving us with:

\begin{equation}
E_{dd} = \left(E_{dd}\right)_{xy} + \left(E_{dd}\right)_z
\label{eq:dipole_split}
\end{equation}
 Allowing us to apply the Dipole-Dipole interaction to the two systems independently.

\subsection{Parameterizing of Unique Hamiltonians and construction of Reduced Configuration Spaces}

We make the assumption that the lowest energy state of the crystal can always be represented as some tile-able state over the infinite lattice. This simply implies periodic boundary conditions for all sums over any finite cell.

It then follows that for a Potts crystal consisting of $N$ sites with magnetic moments defined to be along one of $q$ possible directions, we may define the finite configuration space, ${\mathbb M}$ consisting of $q^N$ elements. Then for any particular state, $M$ living in the configuration space $\mathbb{M}$, we may then define the constants; $C_\alpha\left[ M \right]$, $\left< M \right>_i$, $G_\alpha \left[ M \right]$, and $\left< E_{dd} \right>$ using Eq.\ref{eq:c_alpha} to Eq.\ref{eq:e_avg}.

\begin{align}
C_\alpha\left[ M \right] &= \sum_{\left< i, j\right>_\alpha} \hat{\bf S}_i \cdot \hat{\bf S}_j\label{eq:c_alpha}\\
\left< M \right>_i &= \sum_{j} \hat{\bf S}_j \cdot \hat{\bf e}_i\label{eq:m_i}\\
G_\alpha \left[ M \right] &=  \sum_{\left< i, j\right>_\alpha} \left( \hat{\bf S}_i \times \hat{\bf S}_j \right) \cdot \hat{\bf D}_{ij}\label{eq:g_alpha}\\
\left< E_{dd} \right> &=
\scalebox{0.9}{$\displaystyle
  \sum_{\alpha,\left< i, j\right>_\alpha} \frac{\left( 3 \left( \hat{\bf S}_i \cdot \hat{\bf r}_{ij} \right)   \left( \hat{\bf S}_j \cdot \hat{\bf r}_{ij} \right) - \left( \hat{\bf S}_i \cdot \hat{\bf S}_{j}\right) \right)}{ \left| \vec{\bf r}_{ij} \right|^3}
$}\label{eq:e_avg}
\end{align}

By inserting these definitions into Eq.\ref{sup:eq:hamiltonian_alpha} with the inclusion of DM and Dipole-Dipole Interactions we arrive at a simplified Hamiltonian for a particular state, $M$.

\begin{align}
H_M[J_\alpha,D_\alpha,h_i,g_{dd}] &= -\sum_\alpha \left( J_\alpha C_\alpha + D_\alpha G_\alpha \right) \nonumber\\
&-\sum_i h_i \left< M \right>_i - g_{dd} \left< E_{dd} \right> \label{eq:simplified_ham}
\end{align}

Each Hamiltonian for a state in $\mathbb{M}$ can then be characterized by $2\alpha + 4$ parameters\footnotemark[111]. By collecting only the states which exhibit unique values for $C_\alpha\left[ M \right]$, $G_\alpha \left[ M \right]$, $\left< E_{dd} \right>$ and $\left< M \right>_i$ and therefore by Eq. \ref{eq:simplified_ham} constitute a unique Hamiltonian, we may define a {\it Reduced Configuration Space}, ${\mathbb M}_r \subset {\mathbb M}$.

\footnotetext[111]{It should be noted that because of the result shown in Eq.\ref{eq:dmi_result} that the DM Interaction may be absorbed into the exchange interaction and by immediate consequence $G_\alpha = \sqrt{3} C_\alpha$. Meaning for this particular crystal, we are only effectively parameterizing the Hamiltonian by $\alpha + 4$ parameters.}

As the Reduced Configuration Space, ${\mathbb M}_r$ is a subset of ${\mathbb M}$ consisting of far less elements, it is possible to search over only ${\mathbb M}_r$ to find the state which exhibits the lowest energy for any choice of $J_\alpha$, $D_\alpha$, $h_i$, and $g_{dd}$ in a timely fashion. This method was found to be effective in analyzing the cousin compound, HoAgGe, and is discussed in more detail in Ref.\cite{Gunnar_HoAgGe}.

We do this for both the in-plane non-colinear and the out-of-plane colinear Ising models for all acceptable supercells containing eighteen Dy atoms or less ($N=18$). This results in a Reduced Configuration Space consisting of 235,065 states in-plane and 57,293 states out-of-plane. Of course, the true state observed will always be a superposition of in-plane and out-of-plane states. If we had not separated the Hamiltonian into $H_z$ and $H_{xy}$ we would have found an immense 13,467,579,050 unique Hamiltonians for the four-state system. As searching through thirteen billion states for every parameter combination cannot practically be done in a reasonable time-frame (even for a computer), the reduction of the four-state system into two is a necessity to analyze the Dy lattice in this way. 

\section{Additional Crystal Hamiltonian Interactions}

\subsection{Biquadratic Coupling}

Unlike the Dzyaloshinskii-Moriya and Dipole-Dipole Interactions, the effects of Biquadratic Coupling on the four-state Potts system cannot be divided into the two simpler two-state systems. This comes down purely to the mixing terms in the Biquadratic formula. For any two atoms, $i$, and $j$;

\begin{align}
E_{bq} &= B_{ij} \left( \hat{\mathbf S}_i \cdot \hat{\mathbf S}_j \right)^2\label{eq:biquadratic}\\
&= B_{ij} \left( \mathbf S_i^x\mathbf S_j^x +\mathbf S_i^y\mathbf S_j^y +\mathbf S_i^z\mathbf S_j^z \right)^2\nonumber\\
&= B_{ij} \Bigl[ \left(\mathbf S_i^z\mathbf S_j^z\right)^2 + \left(\mathbf S_i^x\mathbf S_j^x +\mathbf S_i^y\mathbf S_j^y\right)^2\nonumber\\ & \qquad \quad  + 2\mathbf S_i^z\mathbf S_j^z\left(\mathbf S_i^x\mathbf S_j^x +\mathbf S_i^y\mathbf S_j^y\right) \Bigr]\nonumber
\end{align}

While the effects of Biquadratic coupling can be easily quantized into allowed values of $E_{bq}$ by again substituting Eq.\ref{eq:spin_sub}, it cannot be split into the two systems completely independently.

For this reason, we ignored effects of Biquadratic Coupling on the DyAgGe compound as it was incompatible with our analysis technique. We further justify this as $B_{ij}$ is likely much smaller than $J_\alpha$, making Biquadratic effects likely negligible. This assumption was later supported by the accuracy of our results.

\subsection{Symmetric Anisotropic Exchange}

Another interaction which could be included in our analysis is Symmetric Anisotropic Exchange. As the nature of the interaction is symmetric, it can easily be adapted to the new split system. For any two Dy atom $i$ and $j$;

\begin{align}
E_{se} &= \hat{\mathbf S}_i \tilde{T} \hat{\mathbf S}_j \\
&= T_{ij}^x \mathbf S_i^x \mathbf S_j^x + T_{ij}^y \mathbf S_i^y \mathbf S_j^y + T_{ij}^z \mathbf S_i^z \mathbf S_j^z\nonumber
\end{align}
We may then group in-plane and out-of-plane terms as before. Allowing us to define:
\begin{align}
\left(E_{se}\right)_{xy} &= \left( T_{ij}^x \mathbf S_i^x \mathbf S_j^x + T_{ij}^y \mathbf S_i^y \mathbf S_j^y \right)\\
\left(E_{se}\right)_{z} &= \left( T_{ij}^z \mathbf S_i^z \mathbf S_j^z \right)
\end{align}
Of course satisfying:
\begin{equation}
E_{se} = \left(E_{se}\right)_{xy} + \left(E_{se}\right)_{z}
\end{equation}
We may then safely split the effects of Symmetric Anisotropic Exchange between the two systems without affecting the dynamics of the original. $\left(E_{se}\right)_z$ may be easily absorbed into the corresponding Heisenberg exchange parameter, $J_\alpha$. This is not possible for the in plane contribution.

For this reason, in the colinear Ising model, the affects of SAE may be fully absorbed into the Heisenberg Exchange. While this is not true for the non-colinear model, if we assume the affects in the $\hat{\mathbf x}$ and $\hat{\mathbf y}$ are similar, we would expect the affects to be nearly negligible and again absorbable into the Heisenberg Exchange.

While SAE is compatible with our analysis technique, it is not necessary as any affects of SAE would be effectively captured in the Heisenberg interactions and visible in the parameter phase diagrams. As minor adjustments to the Heisenberg parameters do not drastically affect our fitting, we may assume SAE is negligible.

While these considerations partially justify our exclusion of the Biquadratic and Symmetric Anisotropic Exchange affects, the largest reason for their exclusion is the massive computational toll of including them. Including both Biquadratic and SAE would involve paramaterizing our Hamiltonian by an additional $4\alpha$ parameters. By increasing the number of parameters we define unique Hamiltonians by, we massively reduce the degeneracy of the Crystal Configuration Space. This would result in a much larger Reduced Configuration Space. Should the RCS become too large, we would lose much of the efficiency of the method of analysis.

\pagebreak
\section{Results of Analysis}

\subsection{Average Magnetization along Cartesian directions}

Through parameterizing of the Hamiltonian as discussed above, an RCS Search can be performed to find the state exhibiting the lowest energy for the DFT calculated parameters. We generate average magnetization curves for the Dy compound as an external field is applied along the standard Cartesian $\hat{\mathbf x}$, $\hat{\mathbf y}$, and $\hat{\mathbf z}$ directions. These are shown in the main paper and compared against experimental measurements and a Monte Carlo simulation of the system.

\begin{figure}[h!]
\includegraphics[scale=0.5]{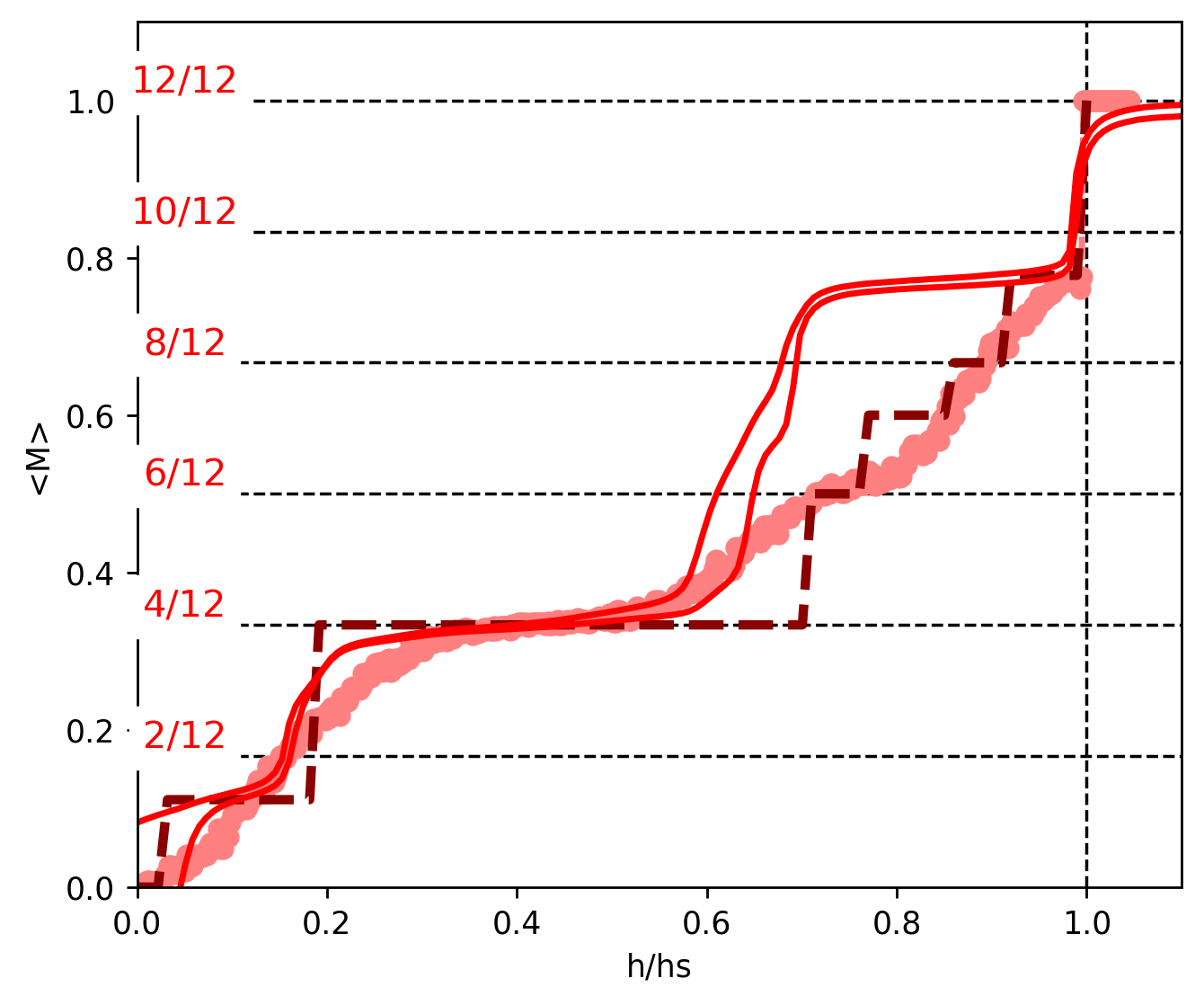}
\centering
\caption{Average magnetization for the Dy atoms as an external magnetic field is applied in the \textcolor{red}{Z} direction. The solid lines represent the experimentally measured magnetization, light dots indicate the magnetization found via MH Monte Carlo, and the dark dashed lines representing magnetization found using an RCS Search.}
\label{fig:magnetization_z}
\end{figure}

\begin{figure}[h!]
\includegraphics[scale=0.25]{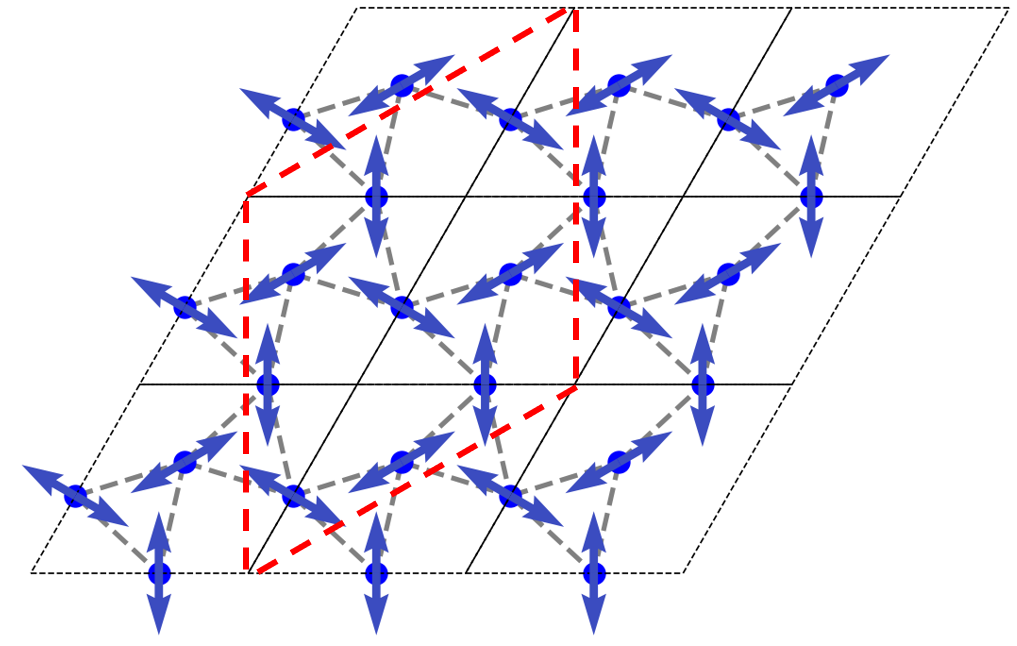}
\centering
\caption{The $\sqrt{2} \times \sqrt{2}$ super cell (also denoted as: [[1 1 0][-1 2 0][0 0 1]]).}
\label{fig:sqrt2}
\end{figure}

\begin{figure}[h!]
\includegraphics[scale=0.45]{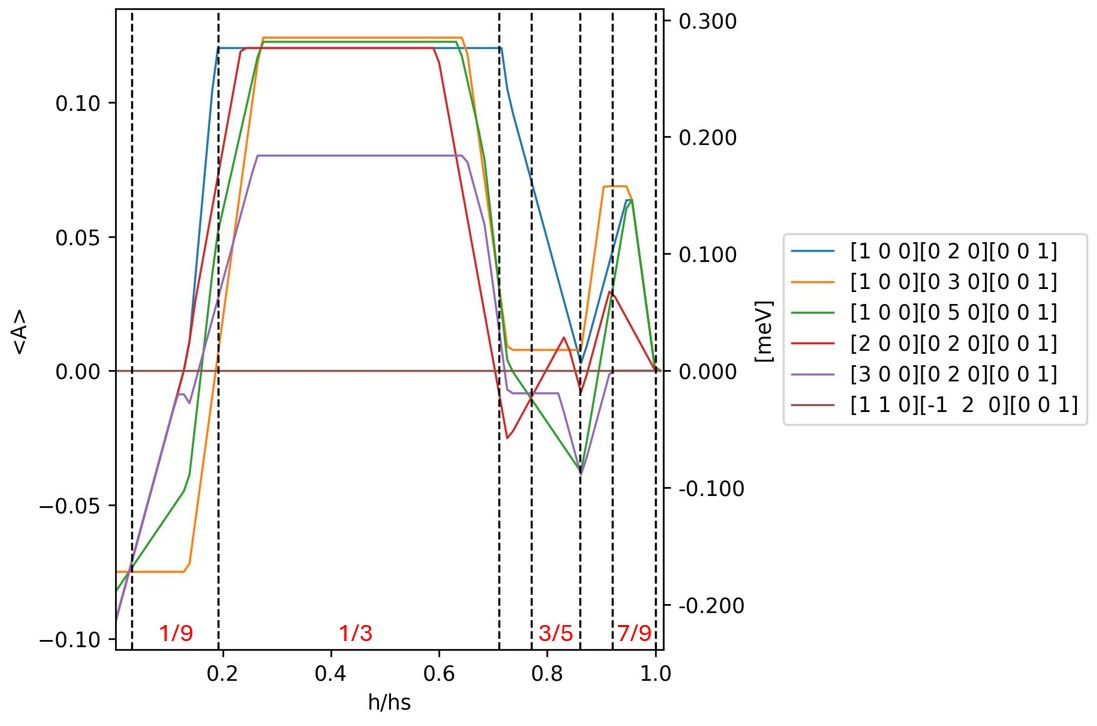}
\centering
\caption{Average energy per Dy atom for differently shaped super cells (labeled in the legend by their respective transformation matrices from the unit cell) as an external field is applied in the $\hat{\mathbf z}$ direction compared against the $\sqrt{2} \times \sqrt{2}$ super cell. Vertical dotted lines indicating magnetic phase transitions.}
\label{fig:energy_specific_cell_z}
\end{figure}

\begin{figure}[h!]
\includegraphics[scale=0.45]{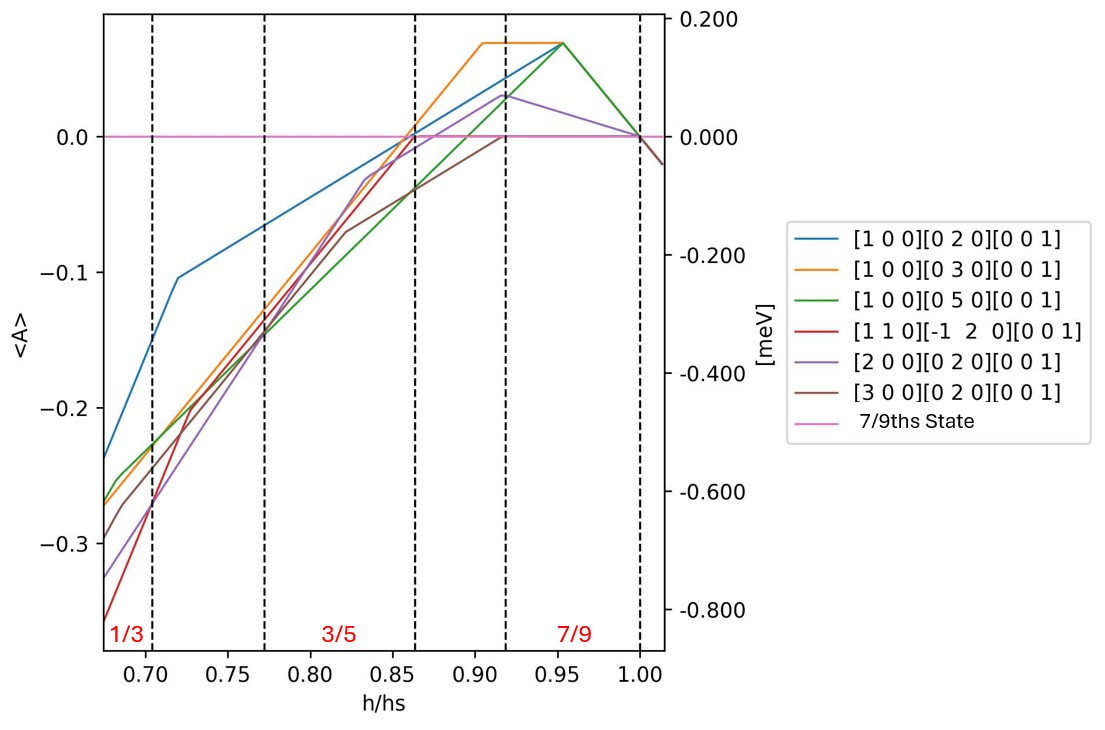}
\centering
\caption{Average energy per Dy atom for differently shaped super cells (labeled in the legend by their respective transformation matrices from the unit cell) as an external field is applied in the $\hat{\mathbf z}$ direction compared against the 7/9ths state. Vertical dotted lines indicating magnetic phase transitions.}
\label{fig:energy_specific_state_z}
\end{figure}

In Fig.\ref{fig:magnetization_z} we can see that in the experiment the crystal shifts from the stable $1/3$rd state to the also stable $7/9$ths state much sooner than is seen using the RCS search or the MC method. By inspection of Fig.\ref{fig:energy_specific_cell_z}, we can see the $1/3$rd and $7/9$ths states are very stable with no competing states being close in energy over their respective ranges for applied external magnetic fields. We believe that because the stable $1/3$rd and $7/9$ths states are constructed on the same super cell (the $\sqrt{2} \times \sqrt{2}$ super cell, Fig.\ref{fig:sqrt2}) and share the same periodicity, that the crystal shifts prematurely to the $7/9$ths state, mostly skipping the interstitial states. These interstitial states are extremely close in energy to the $7/9$ths state, about 0.5 meV at the most apart. This is shown directly in Figures \ref{fig:energy_specific_state_z} and \ref{fig:energy_specific_cell_z} where we directly compare the minimum energy representable on differently shaped super cells against the energy of the $7/9$ths state and the minimum energy of the $\sqrt{2} \times \sqrt{2}$ super cell respectfully. These figures directly show the energy difference between the competing states to be within the temperature threshold of the experiment.

As is shown directly in Fig. \ref{fig:energy_specific_cell_z}, between the $1/3$rd and $7/9$ths states at most a competing cell shape is 0.09 meV less than the $\sqrt{3} \times \sqrt{3}$ super cell. This is well within the allowed temperature threshold and explains why these intersitial states may not be esily observed in experiment between the stable 1/3rd and 7/9ths states.

\pagebreak

\section{Parameter Phase Diagrams}

Using the RCS Search method it is very easy to generate phase diagrams by varying any one of the model parameters as an external field is applied. Below are phase diagrams of the various model parameters as a magnetic field is applied in the $\hat{\mathbf x}$, $\hat{\mathbf y}$, and $\hat{\mathbf z}$ directions. All plots are normalized to the saturation field and $J_1$ = 2.294 meV.

\subsection{Phase Diagrams for Fields Applied Along $\hat{\mathbf x}$}

\begin{figure}[h!]
\includegraphics[scale=0.45]{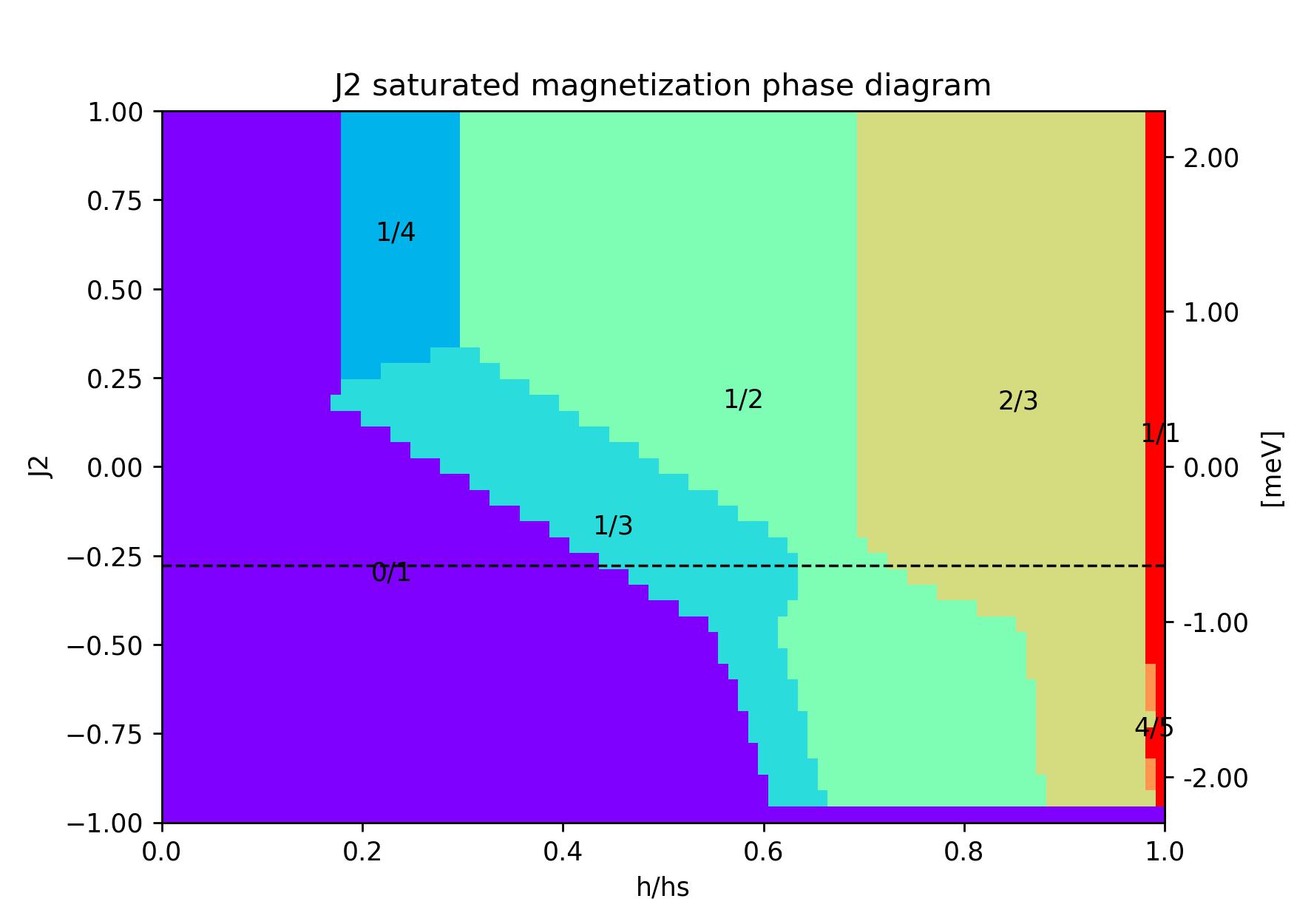}
\includegraphics[scale=0.45]{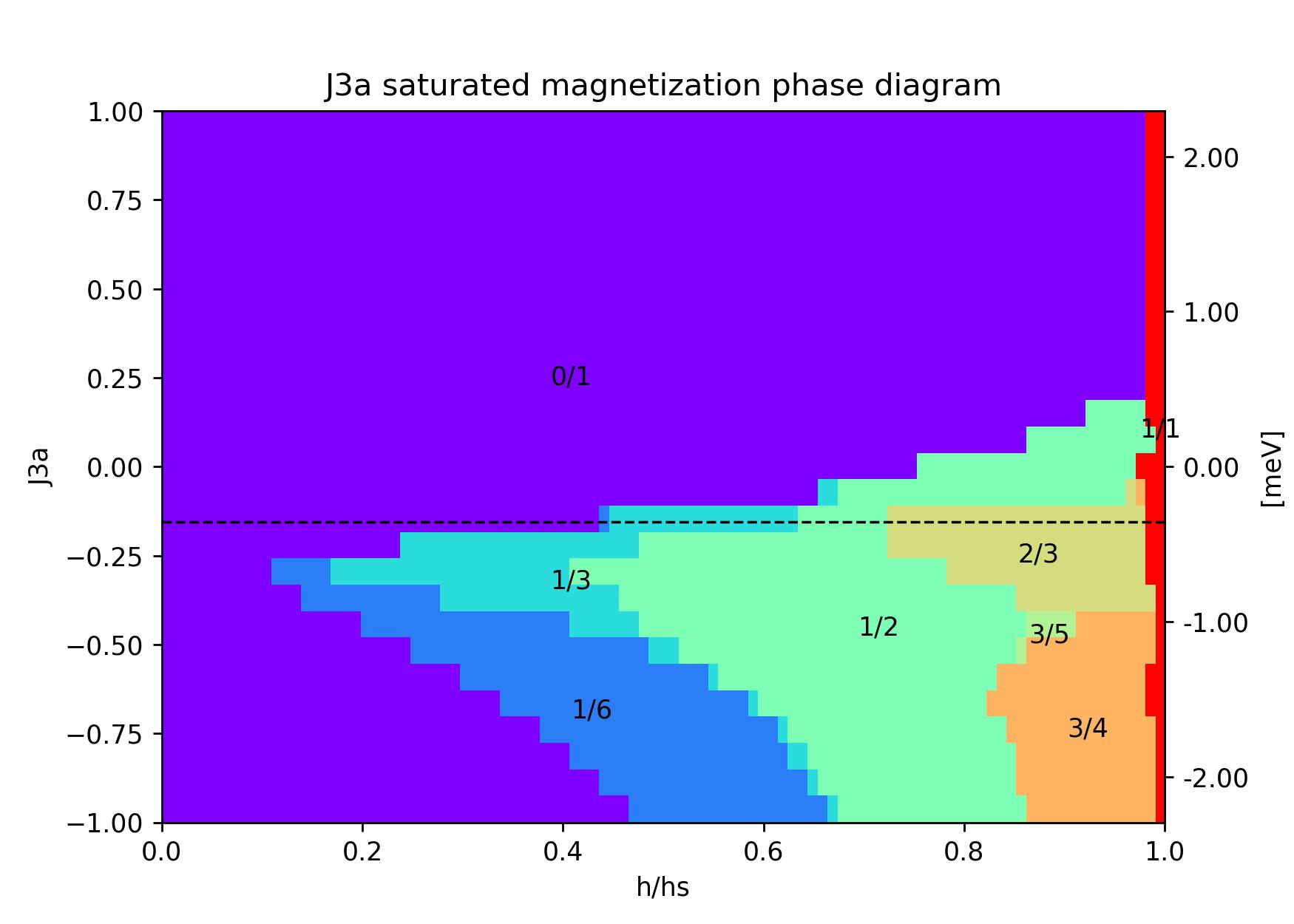}
\includegraphics[scale=0.45]{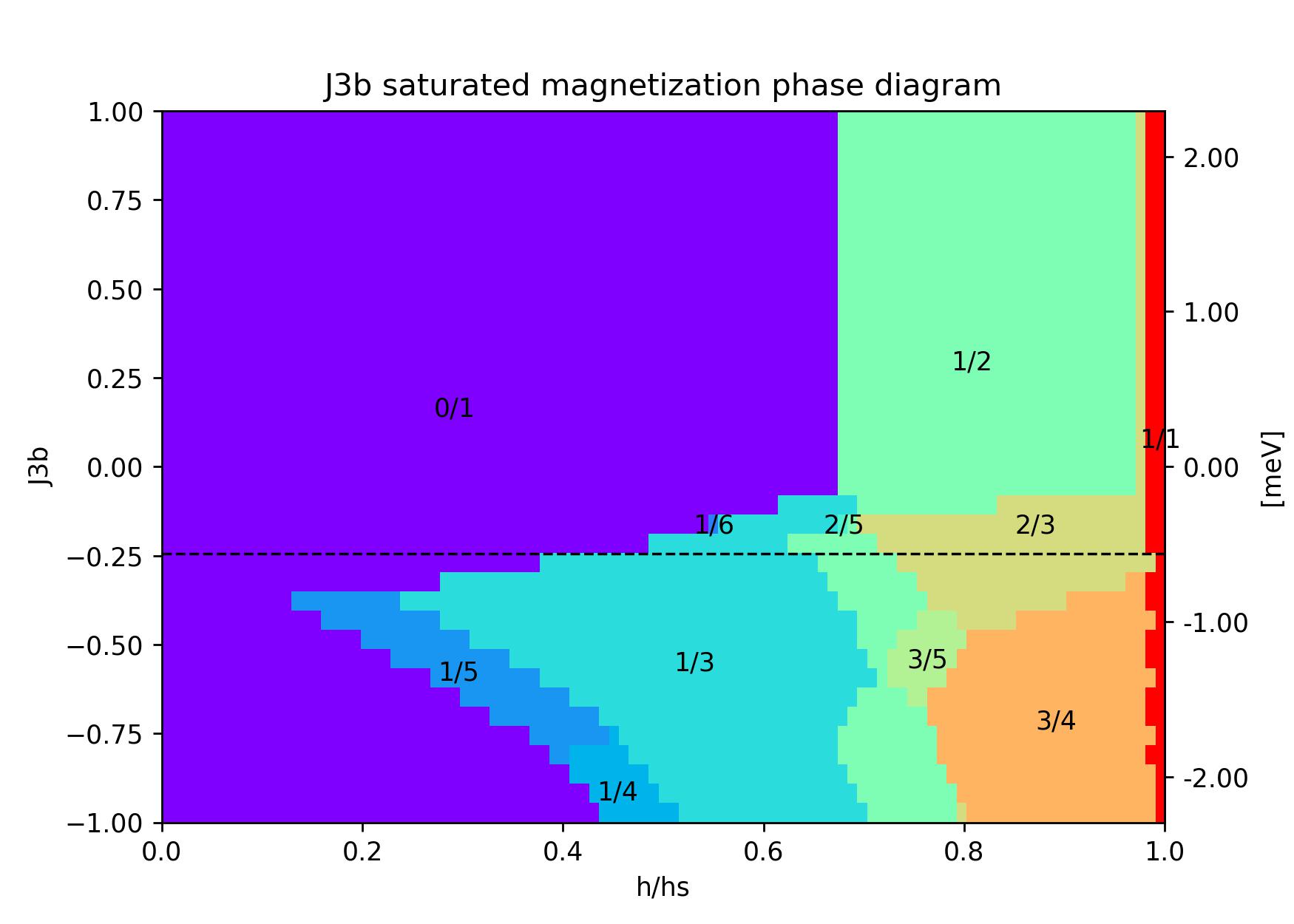}
\includegraphics[scale=0.45]{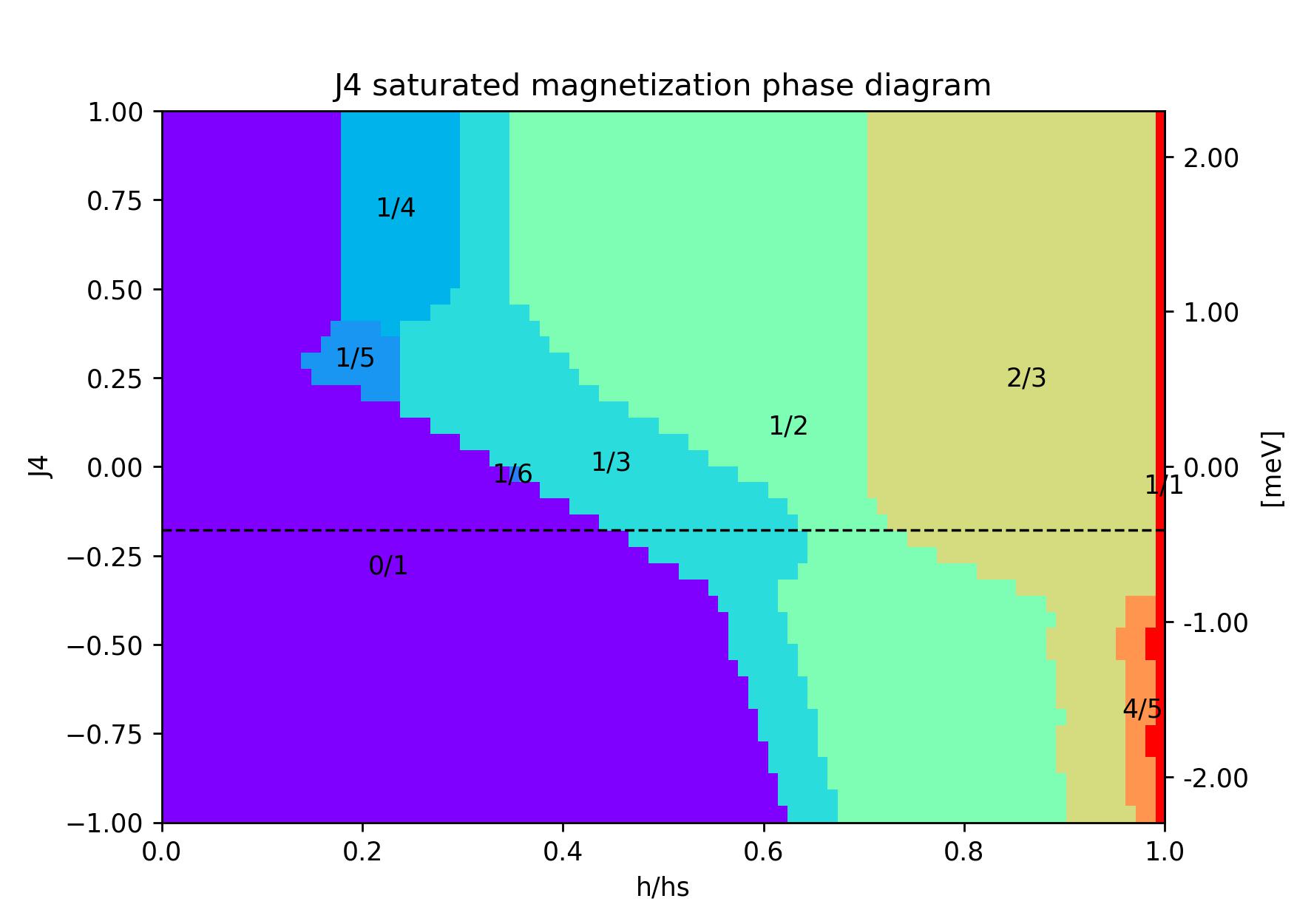}
\includegraphics[scale=0.45]{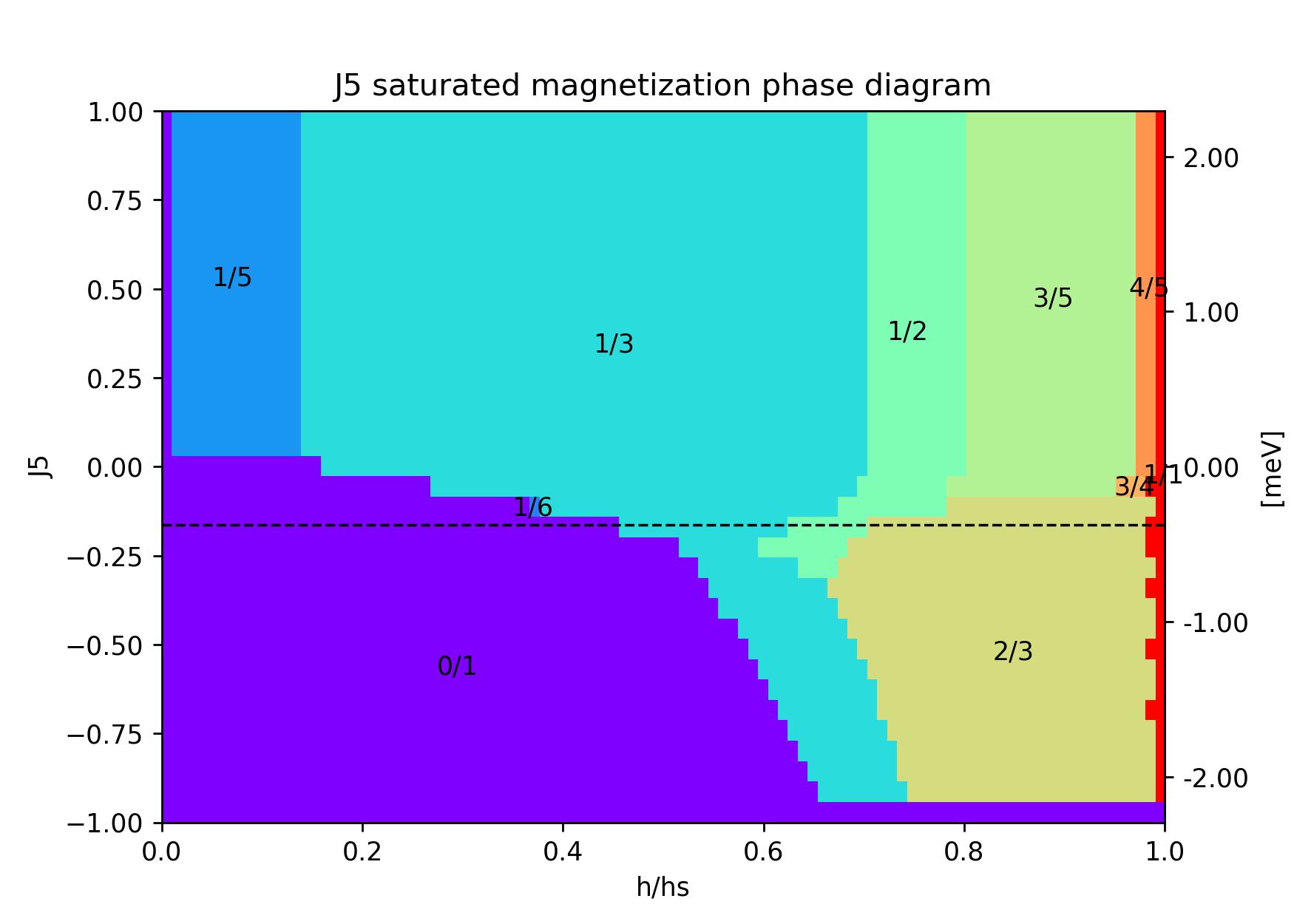}
\includegraphics[scale=0.45]{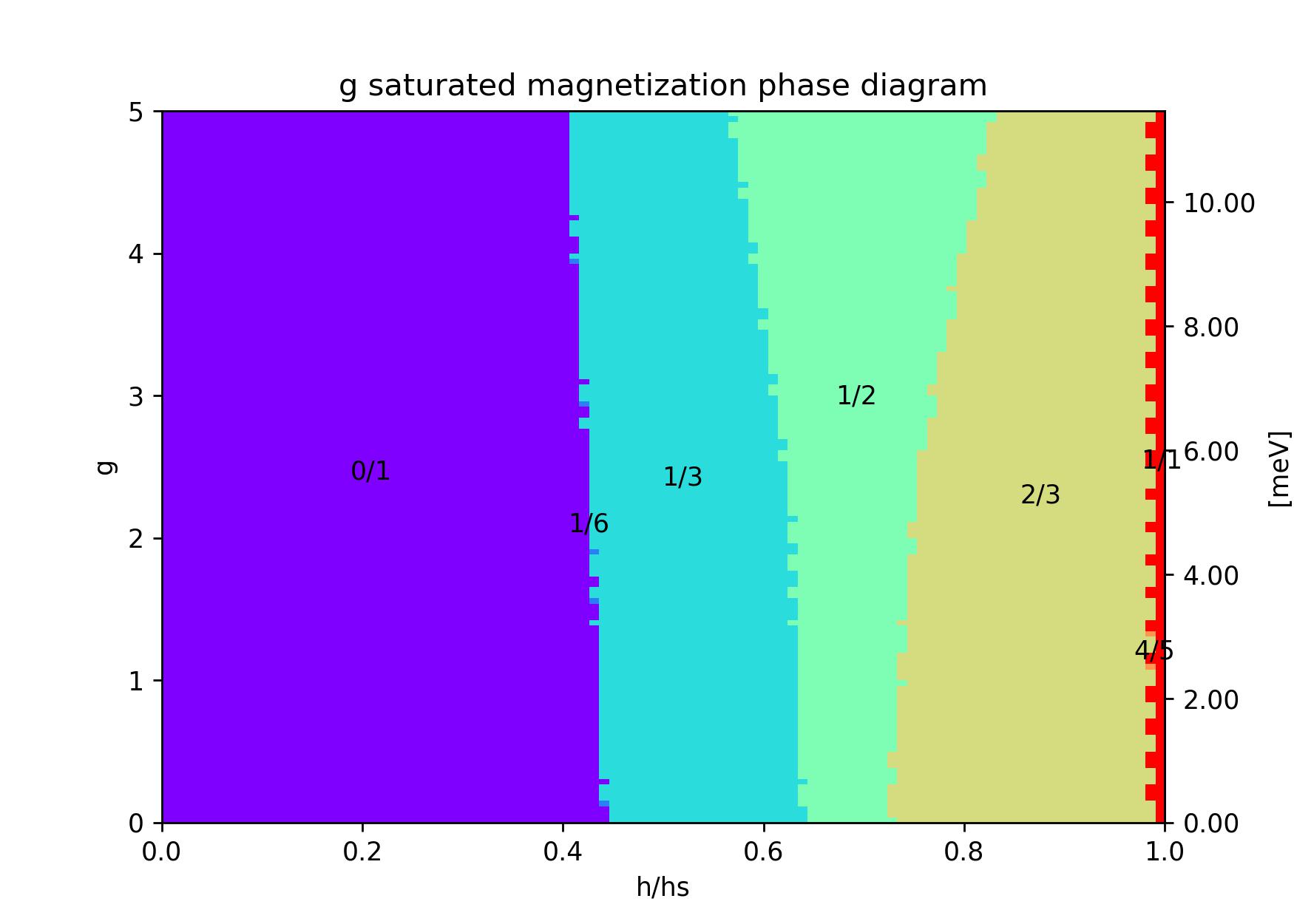}
\centering
\caption{Average magnetization for the Dy atoms as an external magnetic field is applied in the $\hat{\mathbf x}$ direction as Heisenberg exchange parameters are varied for bonds $J_2$ to $J_5$ and $g_{dd}$. Dotted lines indicate the calculated value for the varied parameter.}
\label{fig:phase_x}
\end{figure}

\pagebreak

\subsection{Phase Diagrams for Fields Applied Along $\hat{\mathbf y}$}

\begin{figure}[h!]
\includegraphics[scale=0.45]{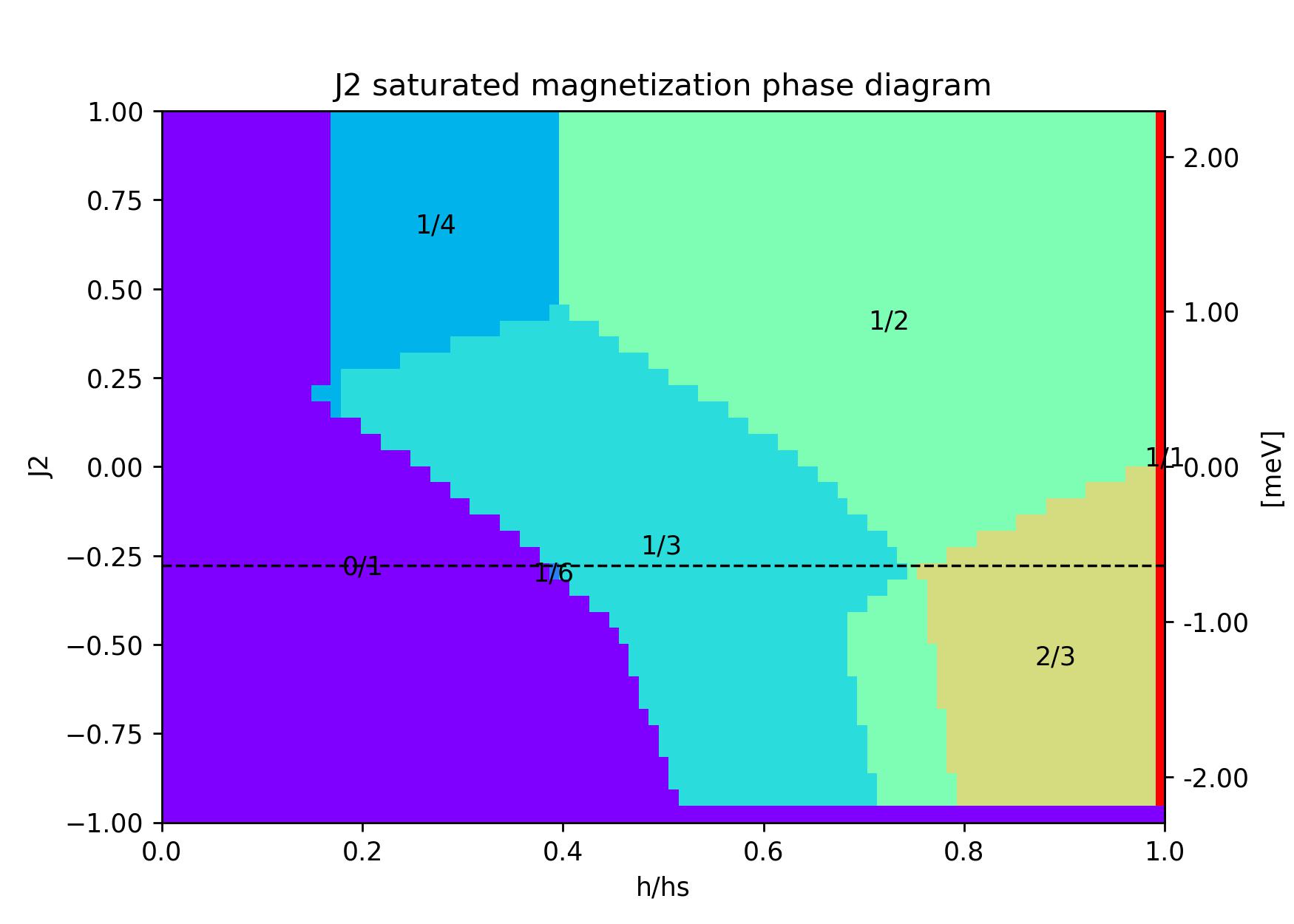}
\includegraphics[scale=0.45]{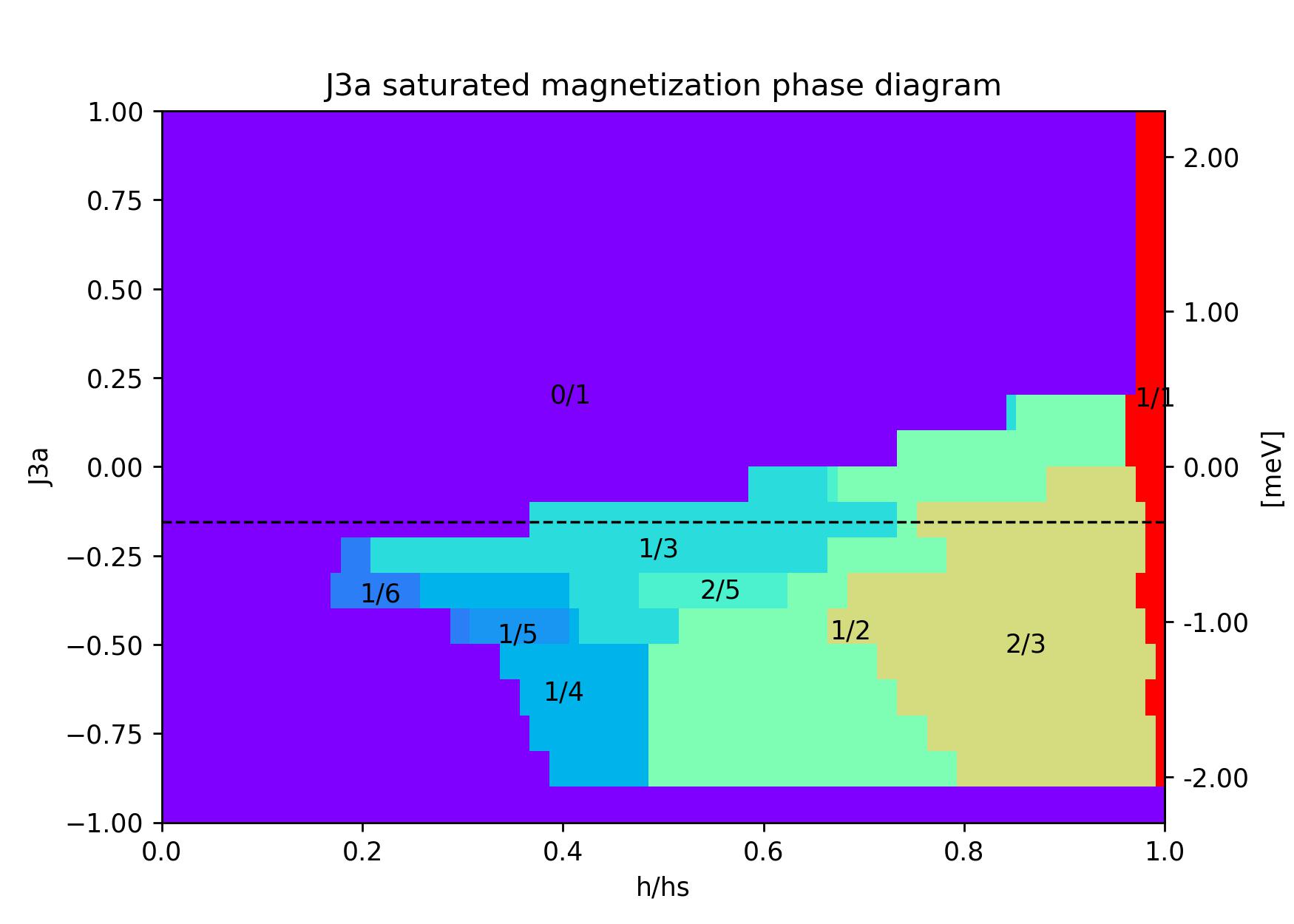}
\includegraphics[scale=0.45]{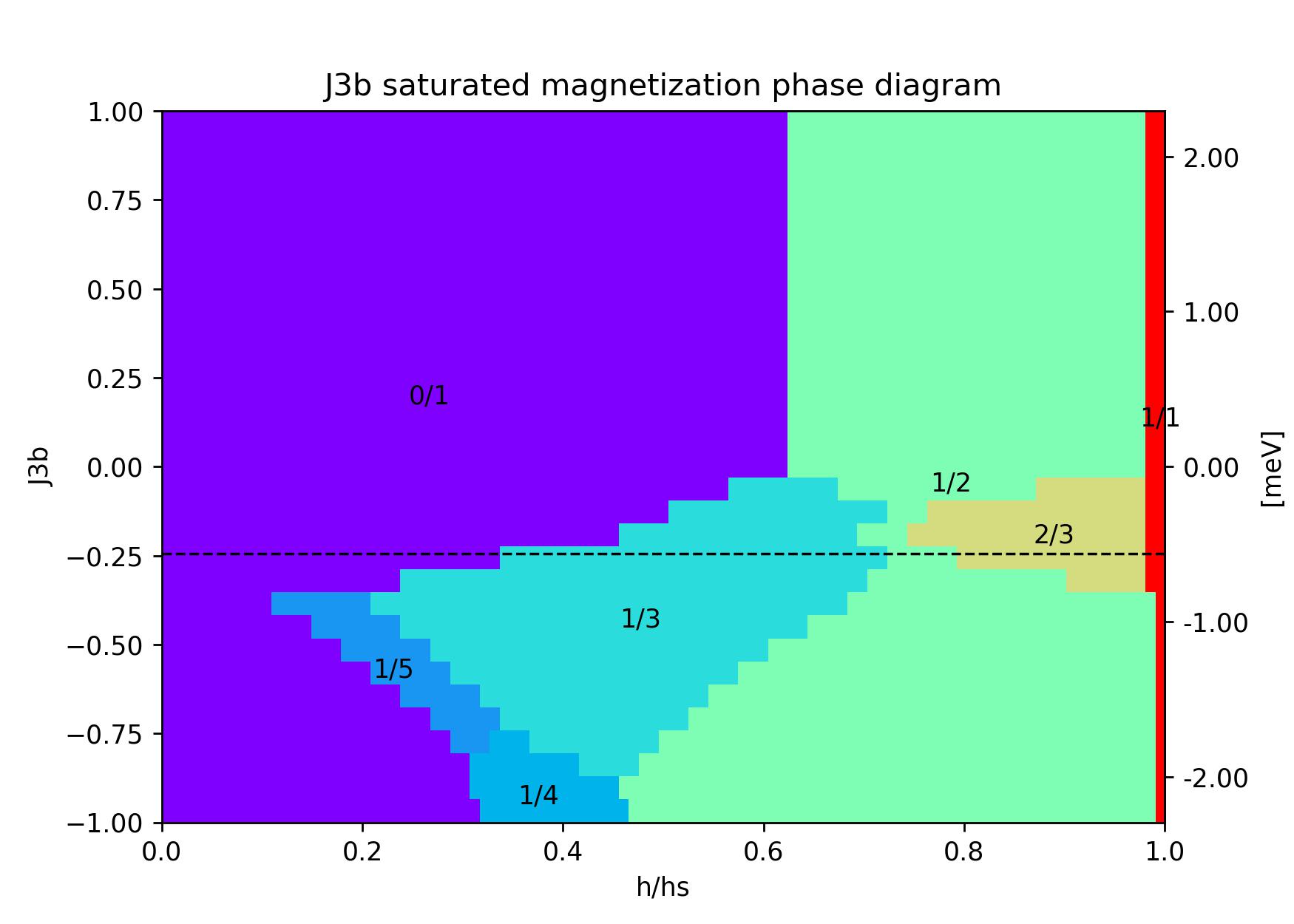}
\includegraphics[scale=0.45]{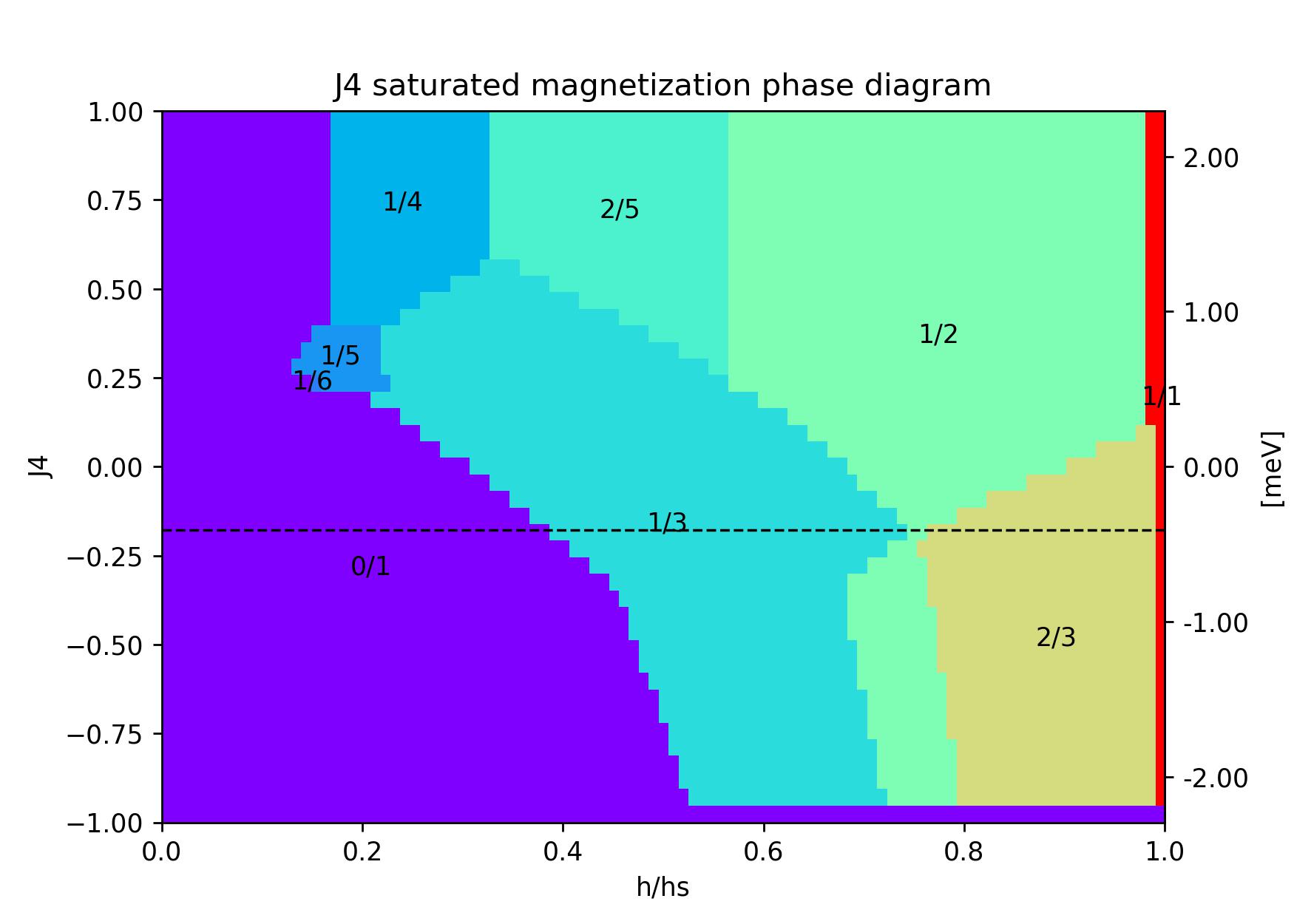}
\includegraphics[scale=0.45]{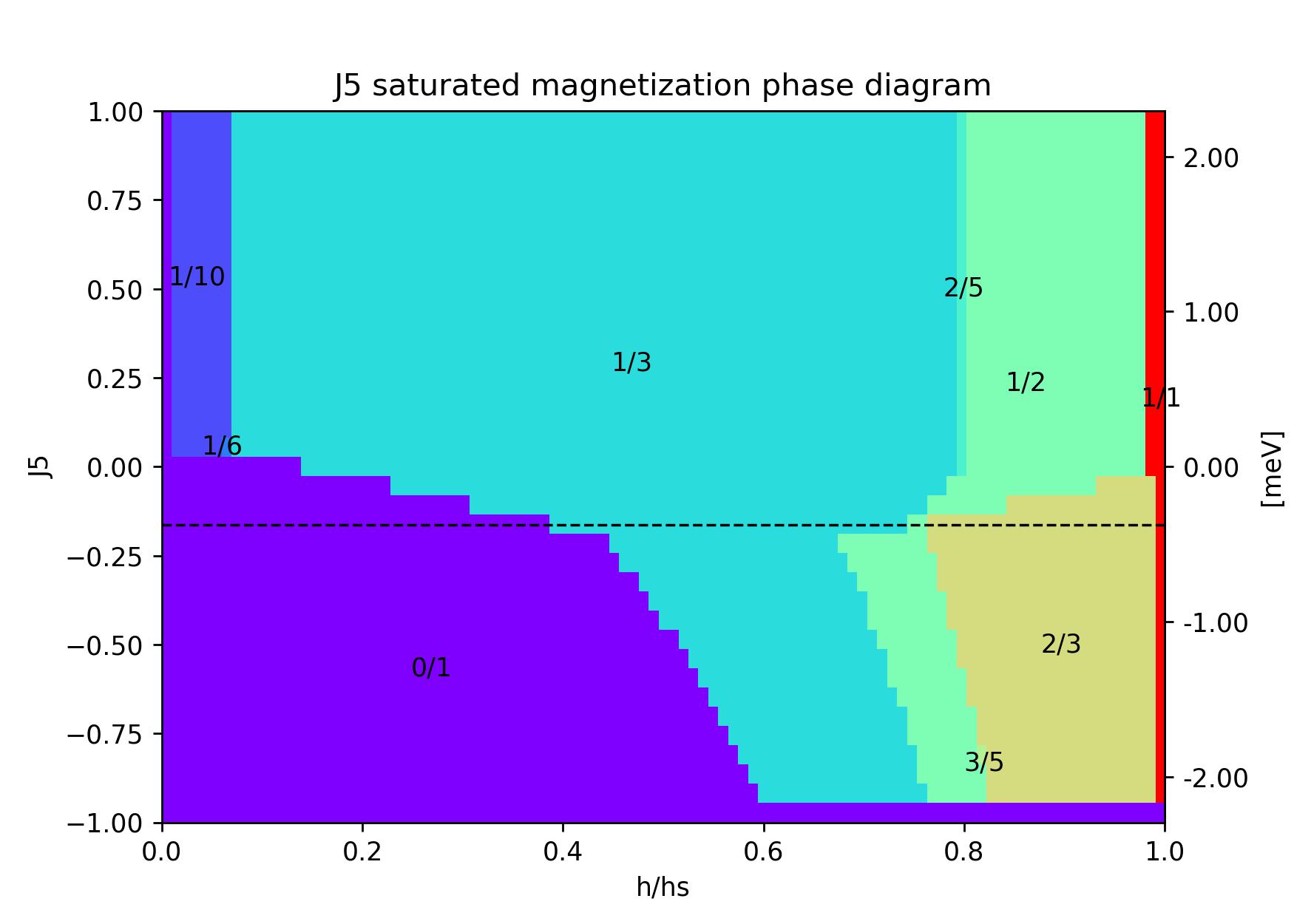}
\includegraphics[scale=0.45]{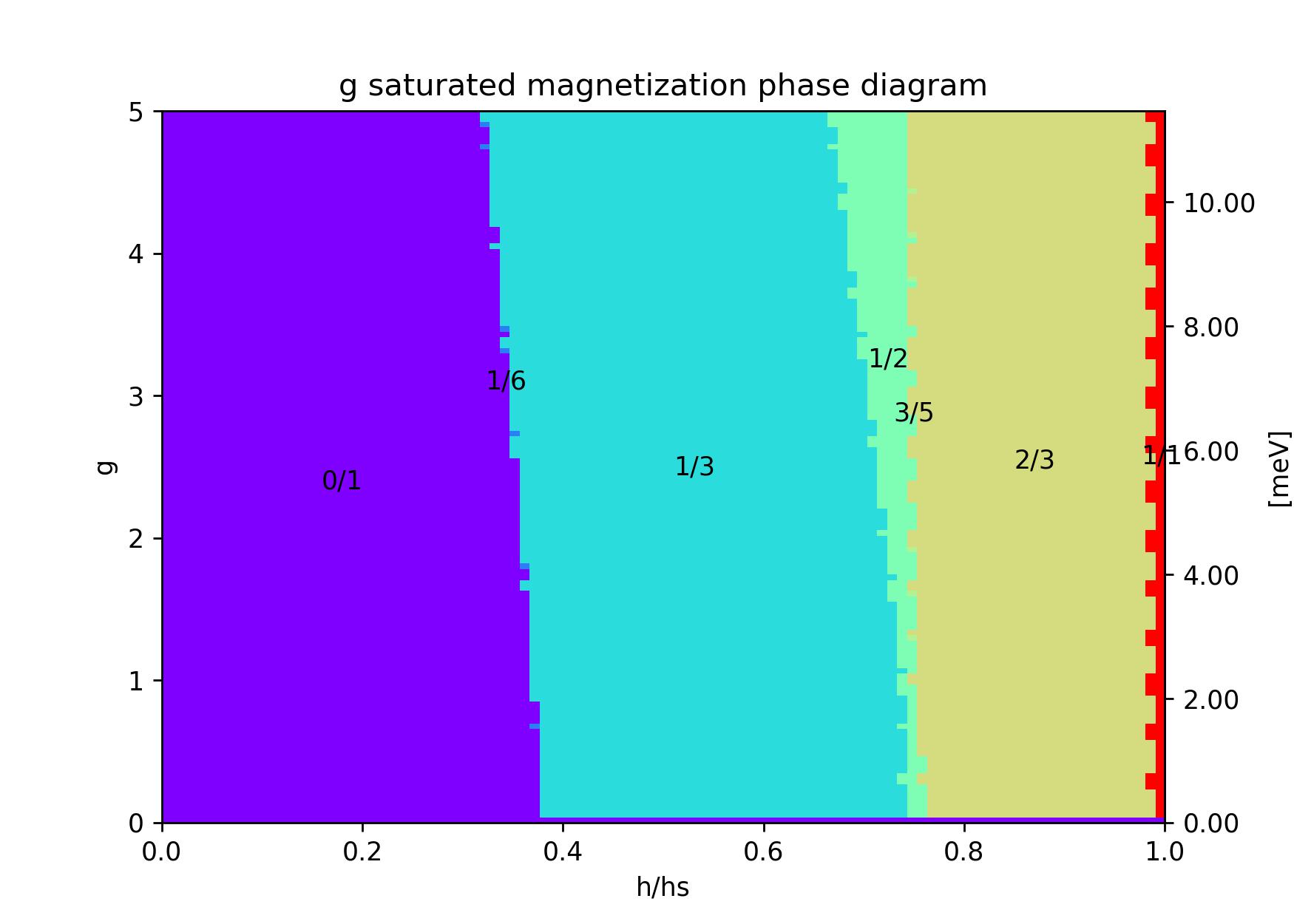}
\centering
\caption{Average magnetization for the Dy atoms as an external magnetic field is applied in the $\hat{\mathbf y}$ direction as Heisenberg exchange parameters are varied for bonds $J_2$ to $J_5$ and $g_{dd}$. Dotted lines indicate the calculated value for the varied parameter.}
\label{fig:phase_y}
\end{figure}

\pagebreak

\subsection{Phase Diagrams for Fields Applied Along $\hat{\mathbf z}$}

\begin{figure}[h!]
\includegraphics[scale=0.45]{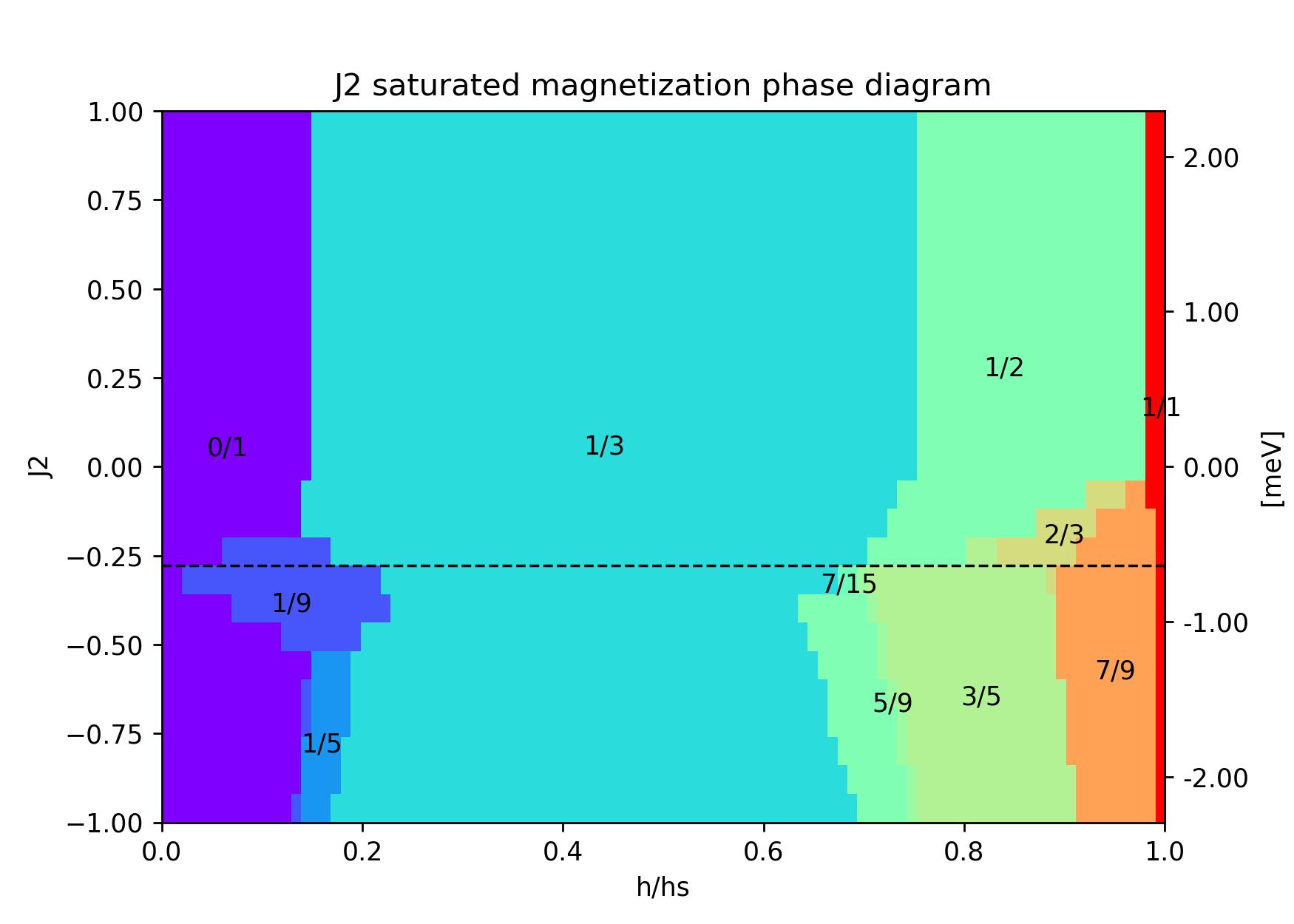}
\includegraphics[scale=0.45]{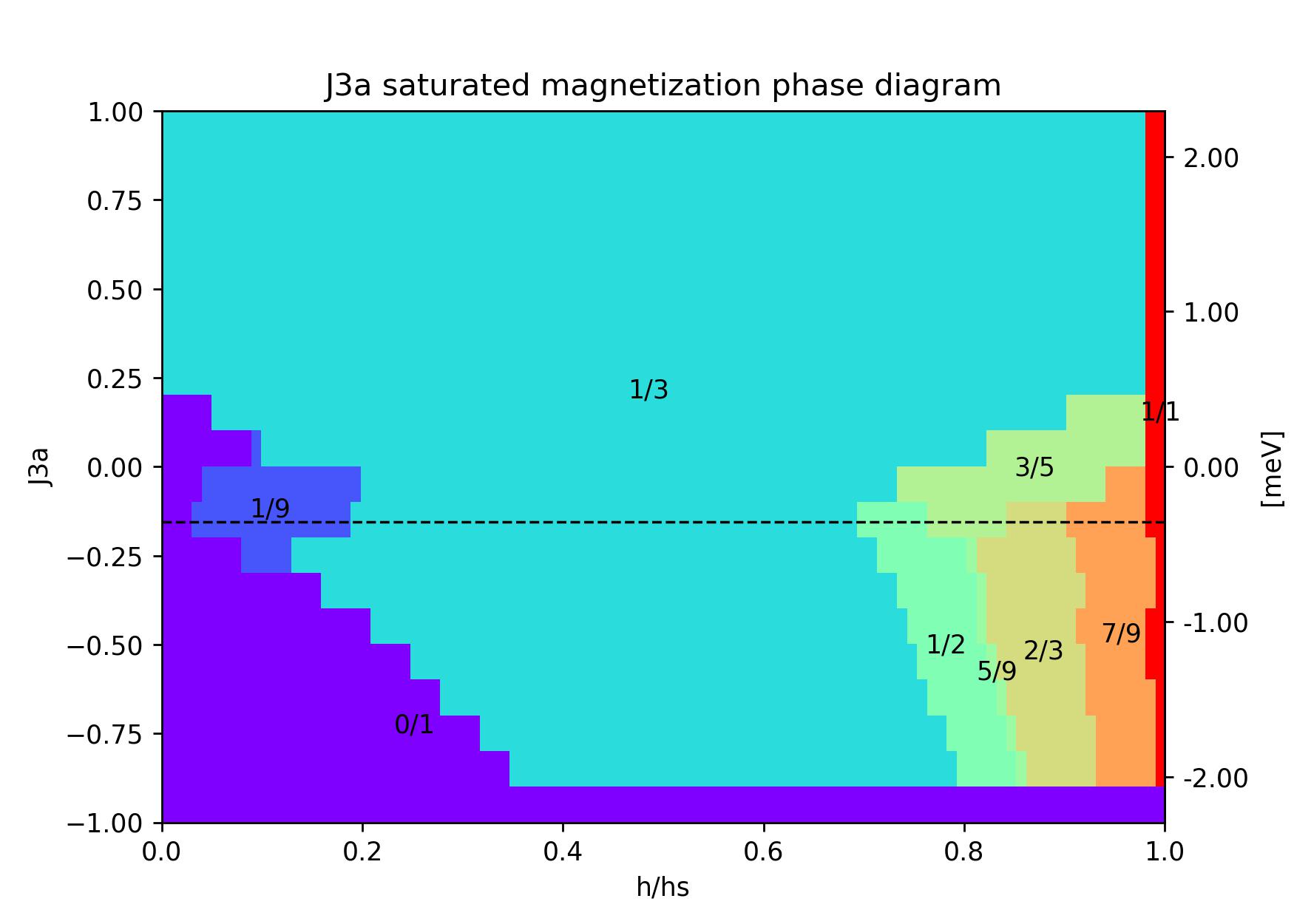}
\includegraphics[scale=0.45]{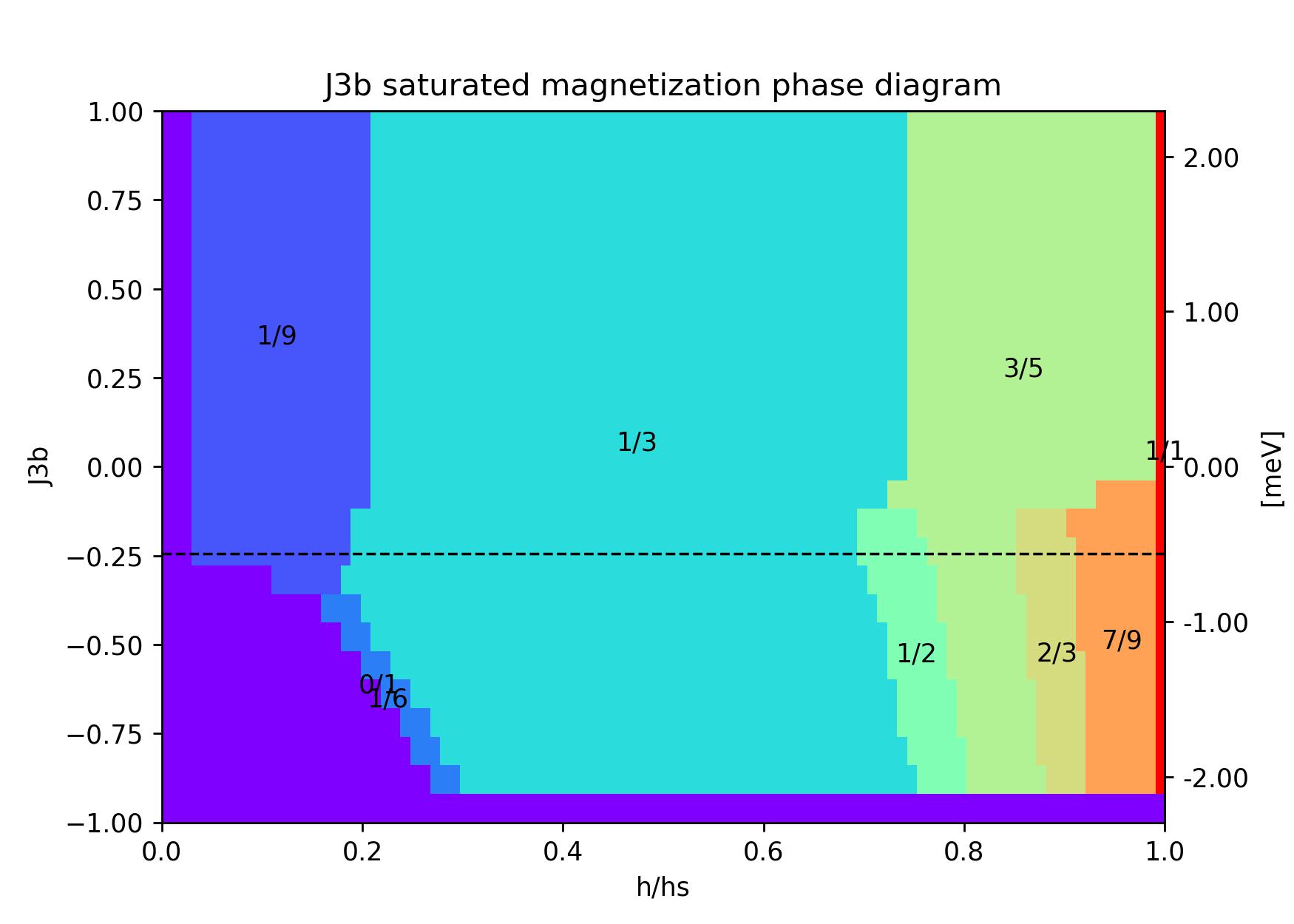}
\includegraphics[scale=0.45]{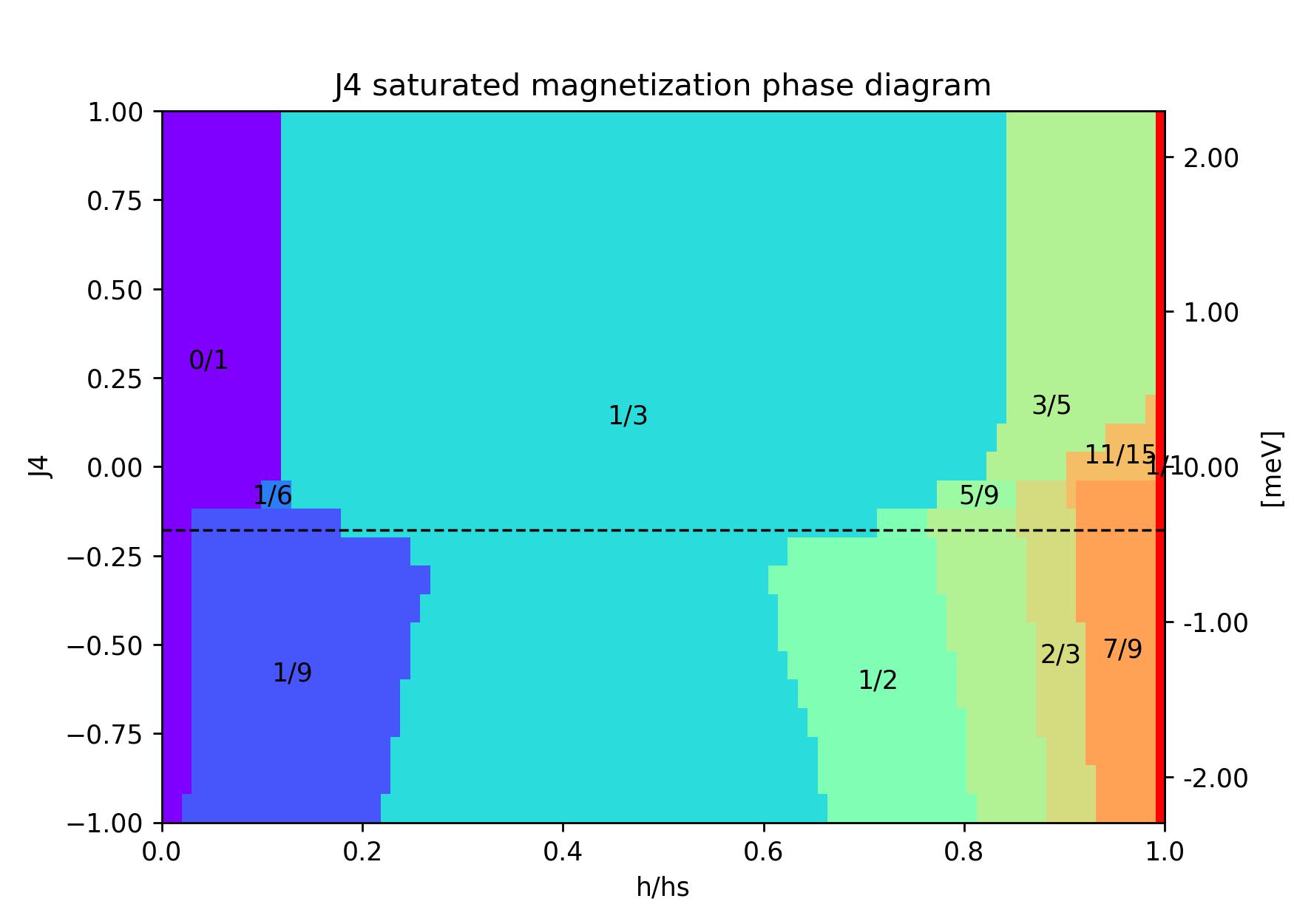}
\includegraphics[scale=0.45]{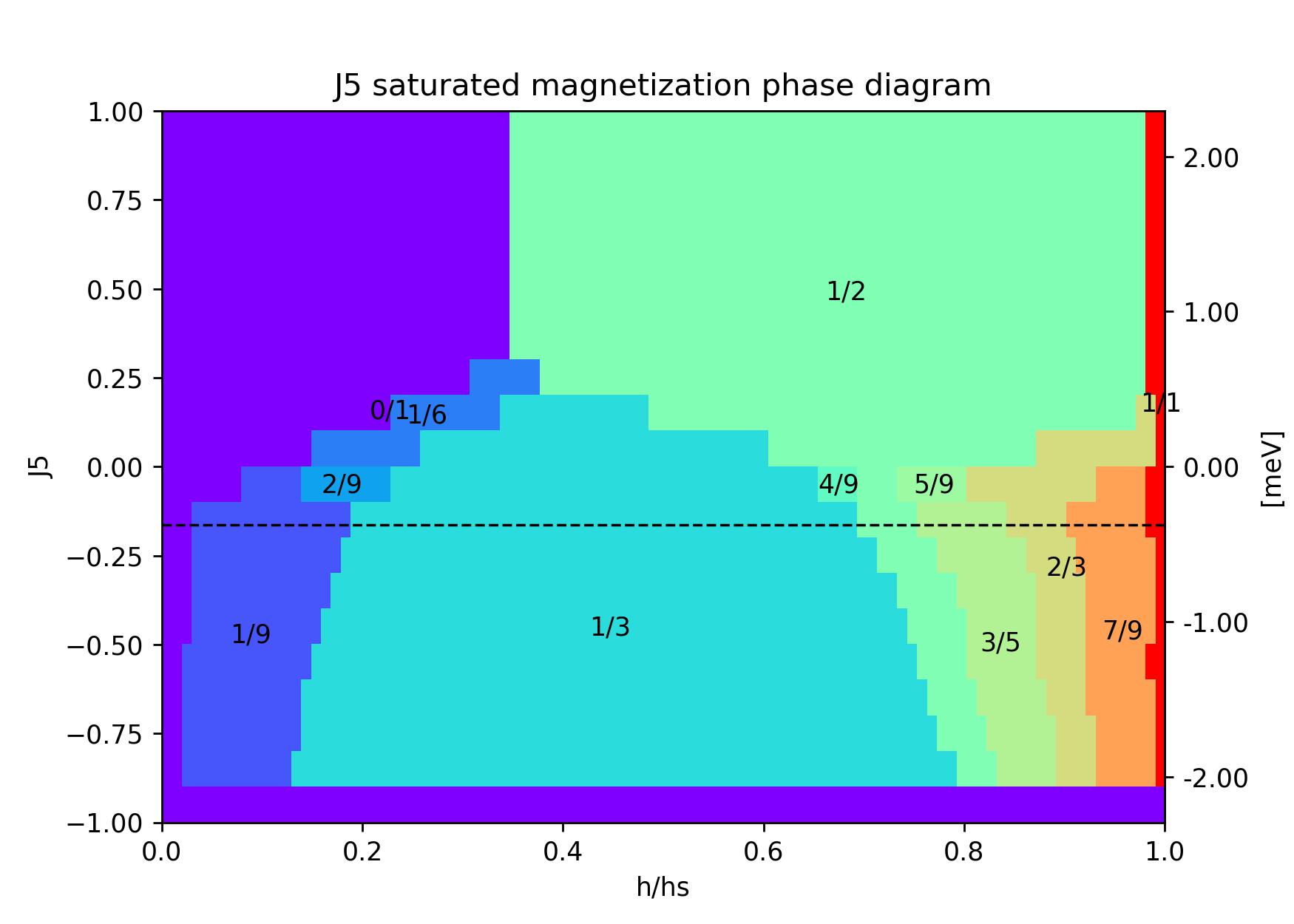}
\includegraphics[scale=0.45]{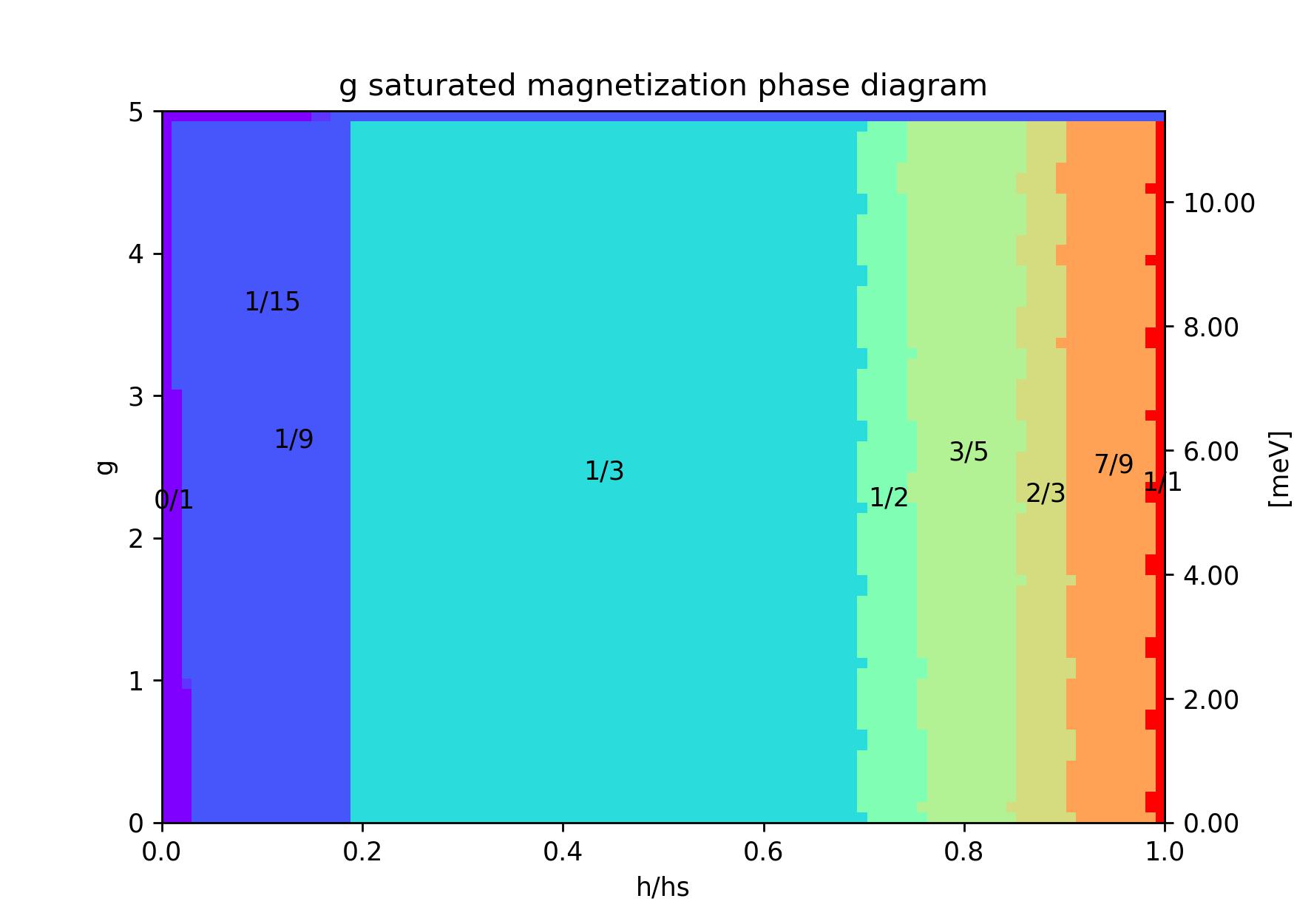}
\centering
\caption{Average magnetization for the Dy atoms as an external magnetic field is applied in the $\hat{\mathbf z}$ direction as Heisenberg exchange parameters are varied for bonds $J_2$ to $J_5$ and $g_{dd}$. Dotted lines indicate the calculated value for the varied parameter.}
\label{fig:phase_z}
\end{figure}

\pagebreak

\section{Experimental Details}

Single crystals of DyAgGe were grown by the flux method considering eutectic point of Ag-Ge. Dy pieces (Thermo Scientific 99.9\%), Ag shots (Thermo Scientific 99.999\%), and Ge pieces (Alfa Aesar 99.9999\%)  were loaded into a 2 mL aluminum oxide crucible in a molar ratio of 0.09:0.6825:0.2275. The crucible was then sealed in a fused silica ampule under vacuum. The sealed ampule was heated to 1175 $^{\circ}$C over 10 hours, homogenized at 1175 $^{\circ}$C for 10 hours, and then cooled to 825 $^{\circ}$C over one week. Once the furnace reached 825 $^{\circ}$C, the tube was centrifuged to separate the crystals in the crucible from the molten flux.  Several well-faceted long crystals up to 100 mg were obtained in the crucible. The crystal structure was verified by Rietveld refinement \cite{mccusker1999rietveld} of a powder X-ray diffraction pattern collected on a pulverized single crystal using a Rigaku Miniflex diffractometer, The Rietveld refinement was performed using the FULLPROF software \cite{rodriguez1993recent}. 

\subsection{Magnetization measurements.}
The magnetization data presented in the manuscript were measured using a Quantum Design (QD) Dynacool Physical Property Measurement System (PPMS).  The AC Measurement System (ACMS II) option was used for the DC magnetization measurements. Sample was oriented with [100], [120] and [001] directions parallel to the applied field for the X, Y and Z directions. Magnetization measurements were verified in more than one crystal for data reproducibility.

\end{document}